\documentclass[english,aps,preprint,superscriptaddress]{revtex4-1}
\usepackage[T1]{fontenc}
\usepackage{amsmath}
\usepackage{amssymb}
\usepackage{graphicx}
\usepackage{esint}

\makeatletter

\usepackage{amsthm}\usepackage{latexsym}\usepackage{bm}\usepackage{amsfonts}

\usepackage{babel}

\usepackage{babel}

\usepackage{hyperref}

\@ifundefined{showcaptionsetup}{}{%
 \PassOptionsToPackage{caption=false}{subfig}}
\usepackage{subfig}
\makeatother

\usepackage{babel}
\begin{document}

\title{Structure-Preserving Geometric Particle-in-Cell Methods for Vlasov-Maxwell
Systems}

\author{Jianyuan Xiao}

\affiliation{School of Physics, University of Science and Technology of China,
Hefei, 230026, China}

\author{Hong Qin}
\email{hongqin@ustc.edu.cn}

\affiliation{School of Physics, University of Science and Technology of China,
Hefei, 230026, China}

\affiliation{Plasma Physics Laboratory, Princeton University, Princeton, NJ 08543,
U.S.A}

\author{Jian Liu}

\affiliation{School of Physics, University of Science and Technology of China,
Hefei, 230026, China}
\begin{abstract}
Recent development of structure-preserving geometric particle-in-cell
(PIC) algorithms for Vlasov-Maxwell systems is summarized. With the
arriving of 100 petaflop and exaflop computing power, it is now possible
to carry out direct simulations of multi-scale plasma dynamics based
on first-principles. However, standard algorithms currently adopted
by the plasma physics community do not possess the long-term accuracy
and fidelity required in these large-scale simulations. This is because
conventional simulation algorithms are based on numerically solving
the underpinning differential (or integro-differential) equations,
and the algorithms used in general do not preserve the geometric and
physical structures of the systems, such as the local energy-momentum
conservation law, the symplectic structure, and the gauge symmetry.
As a consequence, numerical errors accumulate coherently with time
and long-term simulation results are not reliable. To overcome this
difficulty and to hardness the power of exascale computers, a new
generation of structure-preserving geometric PIC algorithms have been
developed. This new generation of algorithms utilizes modern mathematical
techniques, such as discrete manifolds, interpolating differential
forms, and non-canonical symplectic integrators, to ensure gauge symmetry,
space-time symmetry and the conservation of charge, energy-momentum,
and the symplectic structure. These highly desired properties are
difficult to achieve using the conventional PIC algorithms. In addition
to summarizing the recent development and demonstrating practical
implementations, several new results are also presented, including
a structure-preserving geometric relativistic PIC algorithm, the proof
of the correspondence between discrete gauge symmetry and discrete
charge conservation law, and a reformulation of the explicit non-canonical
symplectic algorithm for the discrete Poisson bracket using the variational
approach. Numerical examples are given to verify the advantages of
the structure-preserving geometric PIC algorithms in comparison with
the conventional PIC methods.
\end{abstract}


\maketitle
\global\long\def\EXP{\times10^}  
\global\long\def\rmd{\mathrm{d}}  
\global\long\def\rmc{\mathrm{c}}  
\global\long\def\diag{\textrm{diag}}  
\global\long\def\xs{ \mathbf{x}_{sp}}  
\global\long\def\bfx{\mathbf{x}}  
\global\long\def\bfd{\mathbf{d}}  
\global\long\def\bfp{\mathbf{p}}  
\global\long\def\bfv{\mathbf{v}}  
\global\long\def\bfA{\mathbf{A}}  
\global\long\def\bfY{\mathbf{Y}}  
\global\long\def\bfB{\mathbf{B}}  
\global\long\def\bfS{\mathbf{S}}  
\global\long\def\bfG{\mathbf{G}}  
\global\long\def\bfE{\mathbf{E}}  
\global\long\def\bfM{\mathbf{M}}  
\global\long\def\bfQ{\mathbf{Q}}  
\global\long\def\bfu{\mathbf{u}}  
\global\long\def\bfe{\mathbf{e}}  
\global\long\def\bfzig{\mathbf{r}_{\textrm{zig2}}}  
\global\long\def\bfxzig{\mathbf{r}_{\textrm{xzig}}}  
\global\long\def\bfzzig{\mathbf{r}_{\textrm{zzig}}}  
\global\long\def\xzig{\mathbf{x}_{\textrm{zig}}}  
\global\long\def\yzig{\mathbf{y}_{\textrm{zig}}}  
\global\long\def\zzig{\mathbf{z}_{\textrm{zig}}}  
\global\long\def\zigspmvar{\left( \bfx_{sp,l-1},\bfx_{sp,l},\tau \right)}  
\global\long\def\zigspvar{\left( \bfx_{sp,l},\bfx_{sp,l+1},\tau \right)}  
\global\long\def\rme{\mathrm{e}}  
\global\long\def\rmi{\mathrm{i}}  
\global\long\def\rmq{\mathrm{q}}  
\global\long\def\ope{\omega_{pe}}  
\global\long\def\oce{\omega_{ce}}  
\global\long\def\FIG#1{Fig.~\ref{#1}}  
\global\long\def\TAB#1{Tab.~\ref{#1}}  
\global\long\def\EQ#1{Eq.~(\ref{#1})}  
\global\long\def\SEC#1{Sec.~\ref{#1}}  
\global\long\def\APP#1{Appendix~\ref{#1}}  
\global\long\def\REF#1{Ref.~\cite{#1}}  
\global\long\def\DDELTAT#1{\textrm{Dt}\left(#1\right)}  
\global\long\def\DDELTATA#1{\textrm{Dt}^*\left(#1\right)}  
\global\long\def\GRADD{ {\nabla_{\mathrm{d}}}}  
\global\long\def\CURLD{ {\mathrm{curl_{d}}}}  
\global\long\def\DIVD{ {\mathrm{div_{d}}}}  
\global\long\def\CURLDP{ {\mathrm{curl_{d}}^{*}}}  
\global\long\def\DIVDP{ {\mathrm{div_{d}}^{*}}}  
\global\long\def\cpt{\captionsetup{justification=raggedright }}  
\global\long\def\act{\mathcal{A}}  
\global\long\def\calL{\mathcal{L}}  
\global\long\def\calJ{\mathcal{J}}  
\global\long\def\DELTAA{\left( \bfA_{J,l}-\bfA_{J,l}' \right)}  
\global\long\def\DELTAAL{\left( \bfA_{J,l-1}-\bfA_{J,l-1}' \right)}  
\global\long\def\ADAGGER{\bfA_{J,l}^\dagger}  
\global\long\def\ADAGGERA#1{\bfA_{J,#1}^{x/2}}  
\global\long\def\EDAGGER#1{\bfE_{J,#1}^{x/2}}  
\global\long\def\BDAGGER#1{\bfB_{J,#1}^{x/2}}  
\global\long\def\DDT{\frac{\partial}{\partial t}}  
\global\long\def\DBYDT{\frac{\rmd}{\rmd t}}  
\global\long\def\DBYANY#1{\frac{\partial }{\partial #1}} 
\newcommand{\WZERO}[1]{W_{\sigma_0 I}\left( #1 \right)} 
\newcommand{\WONE}[1]{W_{\sigma_1 J}\left( #1 \right)} 
\newcommand{\WONEJp}[1]{W_{\sigma_1 J'}\left( #1 \right)} 
\newcommand{\WTWO}[1]{W_{\sigma_2 K}\left( #1 \right)} 
\newcommand{\WTHREE}[1]{W_{\sigma_3 L}\left( #1 \right)} 
\newcommand{\bfJ}{\mathbf{J}} 
\global\long\def\MQQ{M_{00}} 
\global\long\def\MDQDQ{M_{11}} 
\global\long\def\MDQQ{M_{01}} 

\section{Introduction}

Particle-in-Cell (PIC) methods \cite{Dawson1983,hockney1988computer,birdsall1991plasma}
have been widely applied since 1960s for simulation studies of plasma
physics and beam physics. PIC algorithms numerically solve the Maxwell
equations for electromagnetic fields and Newton's equation for particle
dynamics, and therefore can simulate a wide range of collective phenomena
in plasmas and charged particle beams. Many PIC algorithms and codes
have been developed \cite{Okuda1972,Cohen1982a,Langdon1983,Lee1983,Cohen1989a,Liewer1989,Friedman1991,eastwood1991virtual,Cary1993,villasenor1992rigorous,Parker1993,Grote1998,Decyk1995,Qin00-084401,Qin00-389,Qiang2000,Chen03-463,Qin01-477,esirkepov2001exact,Vay2002,nieter2004vorpal,Huang2006,Chen2011}.
Why do we need new PIC algorithms? The answer is that standard PIC
algorithms do not have the long-term accuracy and fidelity required
by the large-scale simulations enabled by the upcoming exascale computing
power.

Plasmas and charged particle beams \cite{Davidson01-all} in space
and laboratory naturally contain multi-scale structures and dynamics.
For example, in a typical tokamak device the operation time is about
$10^{10}\sim10^{14}$ times of electron's gyro-period. Obviously,
it is difficult to carry out first-principle based simulations of
these multi-scale structures and dynamics using super computers available
today. Recently, computing power over 100 peta floating point operations
per second \cite{fu2016sunway,Liu2016} has been built, and exascale
computing power is expected to be online in the near future. With
these computing power, simulating multi-scale plasma dynamics directly
using the PIC method becomes possible. However, standard PIC algorithms
currently adopted by the plasma physics community do not possess the
long-term accuracy and fidelity that are needed in the study of multi-scale,
complex dynamics of plasmas. This is because conventional PIC algorithms
are based on numerically solving the underpinning differential (or
integro-differential) equations, and the algorithms used in general
do not preserve the geometric structures of the physical systems,
such as the local energy-momentum conservation law, the symplectic
structure, and the gauge symmetry. As a consequence, numerical errors
accumulate coherently with time and long-term simulation results are
not reliable.

To overcome this serious difficulty and to hardness the power of exascale
computers, a new generation of structure-preserving geometric PIC
algorithms \cite{squire2012geometric,xiao2013variational,xiao2015variational,xiao2015explicit,qin2016canonical,he2016hamiltonian,kraus2017gempic,Morrison2017,xiao2017local}
have been developed. This new generation of algorithms utilizes modern
mathematical techniques, such as discrete manifolds, interpolating
differential forms, and non-canonical symplectic integrators, to ensure
gauge symmetry, space-time symmetry and the conservation of charge,
energy-momentum, and the symplectic structure. These highly desired
properties are extremely difficult to achieve using the conventional
PIC algorithms. The purpose of this paper is to summarize the recent
development of the structure-preserving geometric PIC methods and
to show how to implement such an algorithm for practical purpose.
Important theoretical structures and numerical techniques are discussed
in details. For example, the correspondence between discrete space-time
gauge symmetry and discrete charge conservation law is derived.

Before the technical discussion, we would like to briefly review the
historical development of this active research field. Symplectic integrator
for finite dimensional ordinary differential equations (ODEs) is the
first structure-preserving geometric algorithm systematically studied
in 1980s \cite{Ruth83,Feng85,feng1986difference,feng2010symplectic,Forest90,Channell90,Candy91,Hong02,Tang93,Shang94,Shang99,Sanz-Serna94,Sun2000,marsden2001discrete,hairer2006geometric},
even though the idea may have appeared much earlier \cite{Devogelaere56}.
In a symplectic method, the discrete one-step iteration map preserves
a symplectic 2-form exactly as the exact solution of the Hamiltonian
system does. According to theoretical and practical investigations
\cite{feng1986difference,Sanz-Serna94,Shang99,hairer2006geometric},
symplectic integrators can globally bound the numerical errors of
invariants such as total energy and momentum for all time-steps. For
applications to plasma physics and accelerator physics \cite{Ruth83,Forest90,Cary1993},
charged particle dynamics, expressed in terms of canonical momentum,
admits a canonical symplectic structure and the canonical symplectic
integrators apply straightforwardly \cite{Ruth83,Feng85,feng1986difference,feng2010symplectic,Forest90,Channell90,Candy91,Hong02,Tang93,Shang94,Shang99,Sanz-Serna94,Sun2000,marsden2001discrete,hairer2006geometric}.
Recently, explicit high-order non-canonical symplectic methods for
single charged particle dynamics in the $(\boldsymbol{x},\boldsymbol{v})$
coordinates have been discovered \cite{zhang2016explicit,Wang2016,Zhou2017,Zhang2016,he2017explicit}.
For the single guiding-center dynamics in magnetic field, which admits
an intrinsically non-canonical symplectic structure \cite{little1979,littlejohn1981hamiltonian,littlejohn1983variational},
non-canonical symplectic integrators have been developed \cite{qin2008variational,qin2009variational,zhang2014canonicalization,Ellison2015,Burby2017,Kraus2017,Ellison2018}.
If we relax the symplectic constraint to the weaker condition of volume-preserving,
a series of volume-preserving algorithms for a single charged particle
are available \cite{qin2013boris,zhang2015volume,he2015volume,he2016high,He16-172,Tu2016,Higuera2017}.
It is interesting to note that the familiar Boris algorithm \cite{Boris70}
is volume-preserving but not symplectic \cite{qin2013boris,ellison2015comment}.

In plasma physics and beam physics, conservative systems often have
canonical or non-canonical symplectic structures. But, except for
the single charged particle dynamics discussed above, the symplectic
structures for most important systems are usually infinite dimensional,
such as the Vlasov-Maxwell (VM) systems \cite{morrison1980maxwell,marsden1982hamiltonian,morrison1998hamiltonian,squire,qin2014field},
two-fluid model \cite{spencer1982hamiltonian}, ideal magnetohydrodynamics
(MHD) equations \cite{morrison1980noncanonical}, and gyrokinetic
systems \cite{squire,QinFields,qin2007geometric,burby2014hamiltonian}.
Developing canonical and non-canonical symplectic algorithms for these
systems requires techniques to discretize the infinite dimensional
symplectic structures. In addition, as physical systems defined on
the space-time, these systems also have other important structures
independent from the symplectic structure, such as the gauge symmetry
and charge conservation, the space-time symmetry and energy-momentum
conservation. Developing algorithms that are able to preserve these
physical structures as well presents new challenges.

The first structure-preserving symplectic PIC algorithm for the VM
system was developed by Squire et al. in 2012 \cite{squire2012geometric},
and subsequently by Xiao et al. \cite{xiao2013variational,xiao2015variational}.
This is a variational symplectic algorithm based on the geometric
discretization in both spatial and temporal dimensions of the action
principle for the VM system \cite{Low1958,QinFields,qin2007geometric}.
The resulting discrete time advance conserves a non-canonical symplectic
structure. In 1970s, Lewis proposed to discretize the spatial dimension
of the action principle to reduce the system to a set of ODEs without
addressing the technique of symplectic time integration \cite{lewis1970energy,Lewis1972,evstatiev2013variational,stamm2014variational}.
We note that symplectic time integration is the most crucial part
of the symplectic PIC algorithm for the VM system \cite{squire2012geometric,xiao2013variational,xiao2015variational}.
Of course, Lewis's idea was before the concept of symplectic integrator
was fully developed.

The first non-canonical discrete Poisson bracket for the VM system
was given by Xiao et al. in 2015 \cite{xiao2015explicit} using a
variational method over a discrete space-time manifold with Whitney
forms \cite{whitney1957geometric} that are appropriately generalized
to higher orders. Because the Poisson bracket is non-canonical, it
is nontrivial to develop a time advance algorithm that preserves the
non-canonical symplectic structure. Fortunately and surprisingly,
the explicit high-order splitting time integrator discovered by He
et al. \cite{he2015hamiltonian} for the continuous Morrison-Marsden-Weinstein
bracket \cite{morrison1980maxwell,marsden1982hamiltonian,morrison1998hamiltonian}
works perfectly for the discrete Poisson bracket. It was adopted by
Xiao et al. in Ref. \cite{xiao2015explicit} to develop the first
explicit high-order non-canonical symplectic PIC algorithm for the
VM system. Subsequently, He et al. \cite{he2016hamiltonian} and Kraus
et al. \cite{kraus2017gempic} used the method of finite element discrete
exterior calculus to discretize the Morrison-Marsden-Weinstein bracket
directly. The idea of finite dimensional kinetic system is further
studied by Burby \cite{burby2017finite}.

Complementary to the non-canonical symplectic bracket, there also
exists a canonical bracket for the VM system, which has been discretized
to develop a canonical symplectic PIC algorithm \cite{qin2016canonical}.

For the symplectic PIC algorithms of the VM systems described above,
the space and time are discretized using different methods. The time
advance algorithm in the variational symplectic PIC methods \cite{squire2012geometric,xiao2013variational,xiao2015variational}
is derived from the action discretized in time. Once a temporal discretization
is chosen, the time advance algorithm is determined. This type of
time advance is known as the variational symplectic integrator \cite{marsden2001discrete}.
On the other hand, a splitting method is adopted for the non-canonical
symplectic PIC algorithms \cite{xiao2015explicit,xiao2016explicit,he2016hamiltonian,kraus2017gempic}.
In general, splitting methods are applicable to canonical Hamiltonian
systems only. In fact, universal symplectic method for general non-canonical
systems is not known to exist. However, Crouseilles et al. \cite{crouseilles2015hamiltonian}
proposed a splitting method for the the non-canonical Poisson bracket
of the VM system, which, unfortunately is proven to be incorrect \cite{Qin15JCP}.
The correct splitting algorithm is given by He et al. \cite{he2016hamiltonian}
and then quickly applied to the non-canonical symplectic PIC simulations
\cite{xiao2015explicit,xiao2016explicit,he2016hamiltonian,kraus2017gempic}.

To preserve the differential form structure of the VM system, it is
necessary to select an effective spatial discretization scheme, as
first realized by Squire et al. \cite{squire2012geometric}, who adopted
the method of discrete manifold and Whitney form \cite{whitney1957geometric},
which was generalized to higher orders by Xiao et al. \cite{xiao2015explicit}.
The discrete differential form structure can also be preserved using
finite elements as shown by He et al. \cite{he2016hamiltonian} and
Kraus et al. \cite{kraus2017gempic}. In Ref. \cite{xiao2015explicit},
the discrete non-canonical Poisson bracket is derived from the discrete
Lagrangian, and thus automatically satisfies the Jacobi identity.
In Refs. \cite{he2016hamiltonian} and \cite{kraus2017gempic}, the
discrete bracket is not derived from first-principles, and is confirmed
only through involved calculation, where the properties of the finite
elements compatible with exterior calculus are crucial.

As discussed above, one of the benefits of symplectic algorithms is
the ability to bound for all time-steps the errors of dynamic invariants,
such as the energy and momentum. This is well-known and desirable
for systems with low dimensions. For the symplectic PIC algorithms,
the error on total energy-momentum is bounded. However, the degrees
of freedom of a PIC simulation $D$ is large. One or a few constraints
imposed by the error bound on the total energy-momentum will not bring
a significant improvement over the conventional PIC algorithms. The
symplectic condition requires that the algorithm satisfies $D(2D-1)$
constraints as the exact solution does. Therefore, being symplectic
is much stronger than a few constraints on the total energy-momentum.
It is stronger than imposing local conservation laws at every space-time
grid point, and will result in a significant improvement over the
conventional PIC algorithms. However, we note that discrete local
energy-momentum conservation is a constraint independent from being
symplectic and requires additional geometric structures to be preserved.
Xiao et al. \cite{xiao2017local} showed that the spatially discretized
non-canonical VM system of Ref. \cite{squire2012geometric} indeed
admits a discrete local energy conservation law. The discrete local
momentum conservation law has only been proved from a discrete Maxwell
system by generalizing Noether's theorem for continuous symmetry groups
to discrete symmetry groups \cite{Xiao2017}.

Another important local conservation for PIC algorithms is the conservation
of charge, which is first addressed by Eastwood \cite{eastwood1991virtual},
Villasenor and Buneman \cite{villasenor1992rigorous}, and Esirkepov
\cite{esirkepov2001exact}. For spatially and temporally discretized
systems, Squire et al. \cite{squire2012geometric} pointed out that
the discrete local charge conservation is a direct consequence of
discrete gauge symmetry. In this paper, we will give a general proof
of this fact. Therefore, instead of using special tricks to conserve
local charge \cite{eastwood1991virtual,villasenor1992rigorous,esirkepov2001exact,umeda2003new,pinto2014charge,moon2015exact},
it is much easier to design a Lagrangian discretized in both space
and time that is gauge invariant \cite{squire2012geometric,xiao2013variational,xiao2015explicit,qin2016canonical,he2016hamiltonian,kraus2017gempic,xiao2017local},
and automatically guarantees the local charge conservation.

In this paper, we focus on the structure-preserving geometric PIC
algorithms for the VM system, in which the electromagnetic field is
discretized in a space-time grid. The algorithms apply to the Vlasov-Poisson
system as well. However, for the Vlasov-Poisson system, there exists
another approach that is canonically symplectic, pioneered by Cary
and Doxas \cite{Cary1993}. In this approach, the electrostatic potential
is explicitly solved and expressed in terms of particles' positions,
and the Vlasov-Poisson system is transformed into a Hamiltonian system
only in terms of particles' positions and momenta. Since there is
no magnetic field, the system is canonically symplectic and can be
solved by the standard symplectic integrators for canonical systems
\cite{Ruth83,Feng85,feng1986difference,feng2010symplectic,Forest90,Channell90,Candy91,Hong02,Tang93,Shang94,Shang99,Sanz-Serna94,Sun2000,marsden2001discrete,hairer2006geometric}.
This approach is recently adopted by Shadwick et al. \cite{shadwick2014variational},
Webb \cite{webb2016spectral}, and Qiang \cite{Qiang2016}. Note that
this method is effective because the Poisson equation can be solved
for the electrostatic potential in terms of particles' positions by
inverting the Laplacian operator. For the VM system, the time-dependent
electromagnetic field can not be expressed in terms of particles'
positions and momenta, and the electromagnetic field needs to be treated
as an independent dynamic component of the system. 

Recently, structure-preserving geometric algorithms have also been
developed for the Schrodinger-Maxwell \cite{Chen2017} system and
the Klein-Gordon-Maxwell system \cite{Shi2016,Shi2018}, which have
important applications in high-energy-density physics. Another noteworthy
recent development is the metriplectic particle-in-cell integrators
for the Landau collision operator \cite{Hirvijoki2017}.

This paper is organized as follows. In Section II, an introduction
to the symplectic integrator is given. Section III presents two simple
symplectic PIC examples. Advanced structure-preserving geometric PIC
schemes, including a structure-preserving geometric relativistic PIC
algorithm, are given in Section IV. This section also includes a proof
of the correspondence between discrete gauge symmetry and discrete
charge conservation law, a reformulation of the explicit non-canonical
symplectic algorithm for the discrete Poisson bracket using the variational
approach, some numerical results and comparison with the conventional
Boris-Yee PIC method. 

\section{Symplectic integrators}

The idea of symplectic integrators is to construct discrete one-step
iteration maps that preserve the symplectic two-form associated with
the original Hamiltonian systems, just as the exact solution maps
do. In this section, we briefly describe what this means and how to
achieve it. An elementary example is presented to demonstrate the
significance of the symplectic method. For more systematic studies
and examples, Refs.~ \cite{Ruth83,Feng85,feng1986difference,feng2010symplectic,Forest90,Channell90,Candy91,Hong02,Tang93,Shang94,Shang99,Sanz-Serna94,Sun2000,marsden2001discrete,hairer2006geometric}
are recommended.

\subsection{Hamiltonian systems and the symplectic structure }

Hamiltonian system plays an important role in physics. For a canonical
Hamiltonian system with $N$ generalized momenta and $N$ generalized
coordinates, the dynamics of the system is governed by the canonical
Hamiltonian equation
\begin{eqnarray}
\dot{p}_{i} & = & -\frac{\partial H}{\partial q_{i}}~,\\
\dot{q}_{i} & = & \frac{\partial H}{\partial p_{i}}~,
\end{eqnarray}
where $H=H(p_{i},q_{i},t)$ is the Hamiltonian function, $p_{i}$
and $q_{i}$ are generalized momenta and coordinates. This set of
equations can be written in the matrix form,
\begin{eqnarray}
\dot{z} & = & J^{-1}\nabla H(z)~,\label{EqnHamiJ}
\end{eqnarray}
where 
\begin{eqnarray}
J=\left[\begin{array}{cc}
0 & I\\
-I & 0
\end{array}\right]~
\end{eqnarray}
is the canonical symplectic matrix, $I$ is the $N\times N$ identity
matrix, and $z=(p_{i},q_{i})^{T}$. The most interesting and useful
property of the solution of the canonical Hamiltonian equation is
the preservation of the canonical symplectic structure, i.e., if $z_{t}\left(z_{0}\right)$
is a solution map of the Hamiltonian system Eq. (\ref{EqnHamiJ}),
then the following relation holds, 
\begin{eqnarray}
\left(\frac{\partial z_{t}}{\partial z_{0}}\right)^{T}J\left(\frac{\partial z_{t}}{\partial z_{0}}\right)=J~.\label{EqnCANSYMPJ}
\end{eqnarray}
Maps with this property are called canonical transformations. A more
generalized system is the non-canonical Hamiltonian system (or Poisson
system), 
\begin{eqnarray}
\dot{z} & = & \left\{ z,H(z)\right\} ~,\label{EqnHamiPoisson}
\end{eqnarray}
where $z$ is a vector in an $n$-dimensional phase space, and $\left\{ \cdot,\cdot\right\} $
is the Poisson bracket defined as follows. Suppose $F$, $G$ and
$H$ are functions of $z$, a Poisson bracket is a map which maps
two functions $F$ and $G$ into a new function $\left\{ F,G\right\} $,
satisfying the following conditions. 
\begin{enumerate}
\item Anticommutativity, i.e., 
\begin{eqnarray}
\left\{ F,G\right\}  & = & -\left\{ G,F\right\} ~;
\end{eqnarray}
\item Bilinearity, i.e., 
\begin{eqnarray}
\left\{ F+G,H\right\}  & = & \left\{ F,H\right\} +\left\{ G,H\right\} ~;
\end{eqnarray}
\item Product rule, i.e., 
\begin{eqnarray}
\left\{ FG,H\right\}  & = & \left\{ F,H\right\} G+F\left\{ G,H\right\} ~;
\end{eqnarray}
\item Jacobi identity, i.e., 
\begin{eqnarray}
\left\{ F,\left\{ G,H\right\} \right\} +\left\{ G,\left\{ H,F\right\} \right\} +\left\{ H,\left\{ F,G\right\} \right\}  & = & 0~.
\end{eqnarray}
\end{enumerate}
The Poisson bracket is not only a product operator that satisfies
bilinearity and anticommutativity, but also a derivative operator
with product rule and Jacobi identity. We can also express the Poisson
bracket in the matrix form, 
\begin{eqnarray}
\left\{ F,G\right\}  & = & \left(\frac{\partial F}{\partial z}\right)^{T}B\left(z\right)\frac{\partial G}{\partial z}~,
\end{eqnarray}
where $B\left(z\right)$ is an antisymmetric matrix and satisfies
\begin{eqnarray}
\sum_{l}^{n}\left(\frac{\partial B_{i,j}\left(z\right)}{\partial z_{l}}B_{l,k}\left(z\right)+\frac{\partial B_{j,k}\left(z\right)}{\partial z_{l}}B_{l,i}\left(z\right)+\frac{\partial B_{k,i}\left(z\right)}{\partial z_{l}}B_{l,j}\left(z\right)\right) & = & 0~.\label{EqnBSATI}
\end{eqnarray}
The equation of motion of the non-canonical Hamiltonian system can
be written as 
\begin{eqnarray}
\dot{z} & = & B(z)\frac{\partial H}{\partial z}~.
\end{eqnarray}
The evolution of the non-canonical Hamiltonian system is similar to
the canonical case, with the structure matrix $B(z)$ replacing the
constant matrix $J^{-1}$. The solution map of the non-canonical Hamiltonian
system $z_{t}(z_{0})$ preserves the non-canonical structure associated
with $B(z)$, 
\begin{eqnarray}
\frac{\partial z}{\partial z_{0}}B(z_{0})\left(\frac{\partial z}{\partial z_{0}}\right)^{T} & = & B(z)~.
\end{eqnarray}
Here, $\partial z/\partial z_{0}$ is the Jacobian matrix of $z_{t}(z_{0})$
with respect to $z_{0}$. Note that any non-canonical Hamiltonian
system can be locally transformed into a canonical Hamiltonian system
according to the Darboux-Lie theorem. However the transformation is
in general complicated for practical purpose and does not exist globally.

\subsection{Building symplectic algorithms}

The first generation of symplectic algorithms by Ruth, Feng, and others
\cite{Ruth83,Feng85,feng1986difference,feng2010symplectic} is constructed
by generating functions. For example, for any $C^{2}$ function $S_{1}\left(P,q\right)$,
the following transformation from $(p,q)$ to $(P,Q),$ 
\begin{eqnarray}
P & = & p-\frac{\partial S_{1}\left(P,q\right)}{\partial q}~,\\
Q & = & q+\frac{\partial S_{1}\left(P,q\right)}{\partial P}~,
\end{eqnarray}
preserves the canonical symplectic structure. If we choose the generating
function as $S_{1}\left(P,q\right)=\Delta tH\left(P,q\right)$, then
the corresponding transformation 
\begin{eqnarray}
\left\{ \begin{array}{ccc}
p_{l+1} & = & p_{l}-\Delta tH_{q}\left(p_{l+1},q_{l}\right)~,\\
q_{l+1} & = & q_{l}+\Delta tH_{p}\left(p_{l+1},q_{l}\right)~,
\end{array}\right.\label{EqnSEULER}
\end{eqnarray}
is a 1st-order symplectic Euler method, where $l$ is the label of
the time step and 
\begin{eqnarray}
H_{p}(p,q) & = & \frac{\partial H(p,q)}{\partial p}~,\\
H_{q}(p,q) & = & \frac{\partial H(p,q)}{\partial q}~.
\end{eqnarray}
Higher order symplectic schemes can be built using more accurate $S_{1}$.
As an example, we can let
\begin{eqnarray}
S_{1}(P,q)=\Delta t\sum_{i=1}^{s}b_{i}H(P_{i},Q_{i})-h^{2}\sum_{i,j=1}^{s}b_{i}\hat{a}_{ij}H_{q}\left(P_{i},Q_{i}\right)^{T}H_{p}\left(P_{j},Q_{j}\right),
\end{eqnarray}
where 
\begin{eqnarray}
P_{i} & = & p-h\sum_{j=1}^{s}\hat{a}_{ij}H_{q}\left(P_{j},Q_{j}\right)~,\\
Q_{i} & = & q+h\sum_{j=1}^{s}a_{ij}H_{p}\left(P_{j},Q_{j}\right)~,
\end{eqnarray}
and for all $i,j$, 
\begin{eqnarray}
b_{i}\hat{a}_{ij}+b_{j}a_{ji}=b_{i}b_{j}~.
\end{eqnarray}
Then the corresponding symplectic scheme is the symplectic partitioned
Runge-Kutta (SPRK) method,
\begin{eqnarray}
P & = & p-h\sum_{i=1}^{s}b_{i}H_{q}\left(P_{i},Q_{i}\right)~,\label{EqnSPRKP}\\
Q & = & q+h\sum_{i=1}^{s}b_{i}H_{p}\left(P_{i},Q_{i}\right)~.\label{EqnSPRKQ}
\end{eqnarray}

Another method to build symplectic integrators is based on discrete
variations developed by Marsden et al. \cite{wendlandt1997mechanical,marsden1998multisymplectic,marsden2001discrete},
which is a direct generalization of the variational principle. The
Lagrangian $L$ of a dynamic system is a function of $(q,\dot{q})$,
where $q$ is the generalized coordinates $q=(q_{1},q_{2},\dots,q_{m})^{T}$.
The action integral is 
\begin{eqnarray}
S[q]=\int_{t_{0}}^{t_{1}}L\left(q,\dot{q},t\right)\rmd t~.\label{EqnCS}
\end{eqnarray}
The equation of motion, or the Euler-Lagrange equation, is obtained
from
\begin{eqnarray}
\frac{\delta S}{\delta q}\delta q=\lim_{\epsilon\rightarrow0}\frac{S[q+\epsilon\delta q]-S[q]}{\epsilon}=0~.\label{EqnVAR}
\end{eqnarray}
Here, $\delta q\left(t\right)$ is an arbitrary function of $t$ which
obeys the boundary condition 
\begin{eqnarray}
\delta q(t_{0})=\delta q(t_{1})=0~.\label{EqnBON}
\end{eqnarray}
Inserting the $S$ in \EQ{EqnCS} into \EQ{EqnVAR} yields, 
\begin{eqnarray}
\frac{\delta S}{\delta q}\delta q & = & \lim_{\epsilon\rightarrow0}\frac{1}{\epsilon}\int_{t_{0}}^{t_{1}}\left(L\left(q+\epsilon\delta q,\dot{q}+\epsilon\delta\dot{q}\right)-L\left(q,\dot{q}\right)\right)\rmd t\\
 & = & \int_{t_{0}}^{t_{1}}\left(\frac{\partial L\left(q,\dot{q},t\right)}{\partial q}\delta q+\frac{\partial L\left(q,\dot{q},t\right)}{\partial\dot{q}}\delta\dot{q}\right)\rmd t\\
 & = & \int_{t_{0}}^{t_{1}}\left(\frac{\partial L\left(q,\dot{q},t\right)}{\partial q}-\frac{\rmd}{\rmd t}\frac{\partial L\left(q,\dot{q},t\right)}{\partial\dot{q}}\right)\delta q\rmd t+\left.\frac{\partial L\left(q,\dot{q},t\right)}{\partial\dot{q}}\delta q\right|_{t_{0}}^{t_{1}}=0~.\label{EqnDSDQORI}
\end{eqnarray}
Using the boundary conditions \eqref{EqnBON} and the arbitrariness
of $\delta q\left(t\right)$, we obtain the Euler-Lagrange equation
as 
\begin{eqnarray}
\frac{\partial L\left(q,\dot{q},t\right)}{\partial q}-\frac{\rmd}{\rmd t}\left(\frac{\partial L\left(q,\dot{q},t\right)}{\partial\dot{q}}\right)=0~.\label{EqnEL}
\end{eqnarray}
Now let us build a variational symplectic integrator (VSI) for \EQ{EqnEL}.
Instead of discretizing \EQ{EqnEL} directly, the idea of VSI is
to discretize the action integral $S[q]$. As an example, let's discuss
the following 1st-order VSI. In this method, the action is discretized
as 
\begin{eqnarray}
S_{d} & = & \sum_{l=0}^{N-1}L_{d}\left(q_{l},q_{l+1}\right)\Delta t~.
\end{eqnarray}
where 
\begin{eqnarray}
L_{d}\left(q_{l},q_{l+1}\right) & = & L\left(q_{l},\frac{q_{l+1}-q_{l}}{\Delta t}\right)~,\label{Eqn1stDL}
\end{eqnarray}
$\Delta t$ is the time step and $N$ is the total number of time
steps. The discrete time advance can be obtained by minimizing the
action with respect to $q_{l}$, 
\begin{eqnarray}
\frac{\partial S_{d}}{\partial q_{l}} & = & 0,\quad1\leq l\leq N-1~.\label{EqnDELQ}
\end{eqnarray}
Note that $L_{d}\left(q_{l},q_{l+1}\right)$ depends only on $q_{l}$
and $q_{l+1}$, so \EQ{EqnDELQ} can be simplified as 
\begin{eqnarray}
\frac{\partial L_{d}\left(q_{l-1},q_{l}\right)}{\partial q_{l}}+\frac{\partial L_{d}\left(q_{l},q_{l+1}\right)}{\partial q_{l}}=0~.\label{EqnDELQ2}
\end{eqnarray}
Equation (\ref{EqnDELQ2}) is called the discrete Euler-Lagrange equation
\cite{marsden1998multisymplectic,marsden2001discrete,hairer2006geometric}.
When the values of $q_{l-1}$ and $q_{l}$ are known, $q_{l+1}$ can
be calculated. It is also possible to build high-order variational
integrators through high-order discrete action integrals. For example,
\begin{eqnarray}
L_{d}\left(l,l+1\right) & = & \Delta t\sum_{i=1}^{s}b_{i}L\left(Q_{i},\dot{Q}_{i}\right)~,\label{EqnLdSPRK}
\end{eqnarray}
where 
\begin{eqnarray}
Q_{i} & = & q_{l}+\Delta t\sum_{j=1}^{s}a_{ij}\dot{Q}_{j}~
\end{eqnarray}
and $\dot{Q}_{i}$ are chosen to extremize \EQ{EqnLdSPRK} under
the constraints 
\begin{eqnarray}
\sum_{i=1}^{s}b_{i} & = & 1~,\nonumber \\
b_{i} & \neq & 0~,\nonumber \\
q_{l+1} & = & q_{l}+\Delta t\sum_{i=1}^{s}b_{i}\dot{Q}_{i}~.\label{EqnCONSTQI}
\end{eqnarray}
From this discrete action, we can derive the following variational
symplectic partitioned Runge-Kutta (VSPRK) integrator, 
\begin{eqnarray}
p_{l+1} & = & p_{l}+\Delta t\sum_{i=1}^{s}b_{i}\dot{P}_{i}~,\\
q_{l+1} & = & q_{l}+\Delta t\sum_{i=1}^{s}b_{i}\dot{Q}_{i}~,\\
P_{i} & = & p_{l}+\Delta t\sum_{j=1}^{s}\hat{a}_{ij}\dot{P}_{j}~,\\
Q_{i} & = & q_{0}+\Delta t\sum_{j=1}^{2}a_{ij}\dot{Q}_{j}~,
\end{eqnarray}
where 
\begin{eqnarray}
\hat{a}_{ij} & = & b_{j}-b_{j}a_{ji}/b_{i}~,\\
\dot{P}_{i} & = & -H_{q}\left(P_{i},Q_{i}\right)~,\\
\dot{Q}_{i} & = & H_{p}\left(P_{i},Q_{i}\right)~,\\
H\left(p,q\right) & = & p^{T}\dot{q}(p,q)-L\left(q,\dot{q}(p,q)\right)~,\label{EqnHMT}\\
p & = & \frac{\partial L}{\partial\dot{q}}\left(q,\dot{q}\right)~.\label{EqnLEGEND}
\end{eqnarray}
When the Legendre transformation specified by \EQ{EqnLEGEND} is
not degenerate, the above VSPRK integrator is equivalent to the previous
canonical SPRK integrator given by Eqs.~(\ref{EqnSPRKP}) and (\ref{EqnSPRKQ}),
and the 1st-order variational integrator is equivalent to the 1st-order
symplectic Euler method.

For non-canonical Hamiltonian systems, universal symplectic integrator
applicable to all non-canonical systems is not known to exist. But
it is possible to construct symplectic integrators for specific non-canonical
structures. Examples include the symplectic guiding center integrators
\cite{qin2008variational,qin2009variational,zhang2014canonicalization,Ellison2015,Burby2017,Kraus2017,Ellison2018}
and symplectic integrators for the non-canonical charged particle
dynamics in the $(\boldsymbol{x},\boldsymbol{v})$ coordinate \cite{zhang2016explicit,Wang2016,Zhou2017,he2017explicit,Zhang2016}.

\subsection{An example: harmonic oscillator}

The simplest Hamiltonian system is the harmonic oscillator, for which
the Lagrangian is 
\begin{eqnarray}
L\left(q,\dot{q}\right) & = & m\frac{\dot{q}^{2}}{2}-kq^{2}~.
\end{eqnarray}
Here $q$ is the displacement of the oscillator, $k$ and $m$ are
positive constants. The equation of motion, or the Euler-Lagrange
equation, is 
\begin{eqnarray}
\ddot{q} & = & -\frac{k}{m}q~.
\end{eqnarray}
The system is conservative, and the total energy is an invariant of
the dynamics. The orbit is periodic without damping.

Let us build a VSI for it. The Lagrangian can be discretized using
the 1st-order forward difference scheme given by \EQ{Eqn1stDL}.
From the discrete Euler-Lagrange equation, we obtain the discrete
iteration relation 
\begin{eqnarray}
q_{l+1} & = & 2q_{l}-q_{l-1}-\Delta t^{2}\frac{k}{m}q_{l}~.
\end{eqnarray}

\begin{figure}
\subfloat[Relation of $q$ and $\dot{q}$.]{\includegraphics[width=0.49\textwidth]{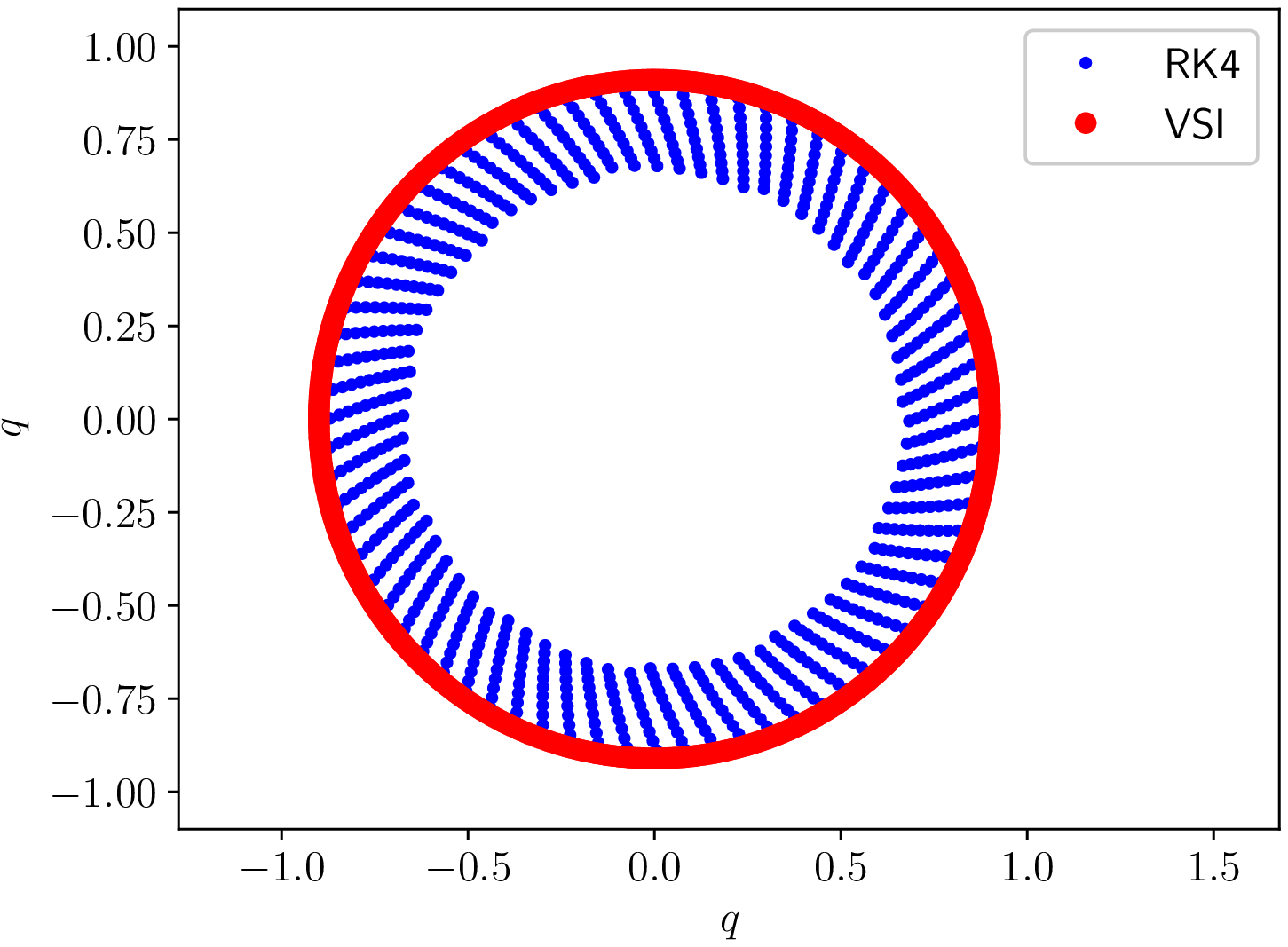}

}\subfloat[Total energy.]{\includegraphics[width=0.49\textwidth]{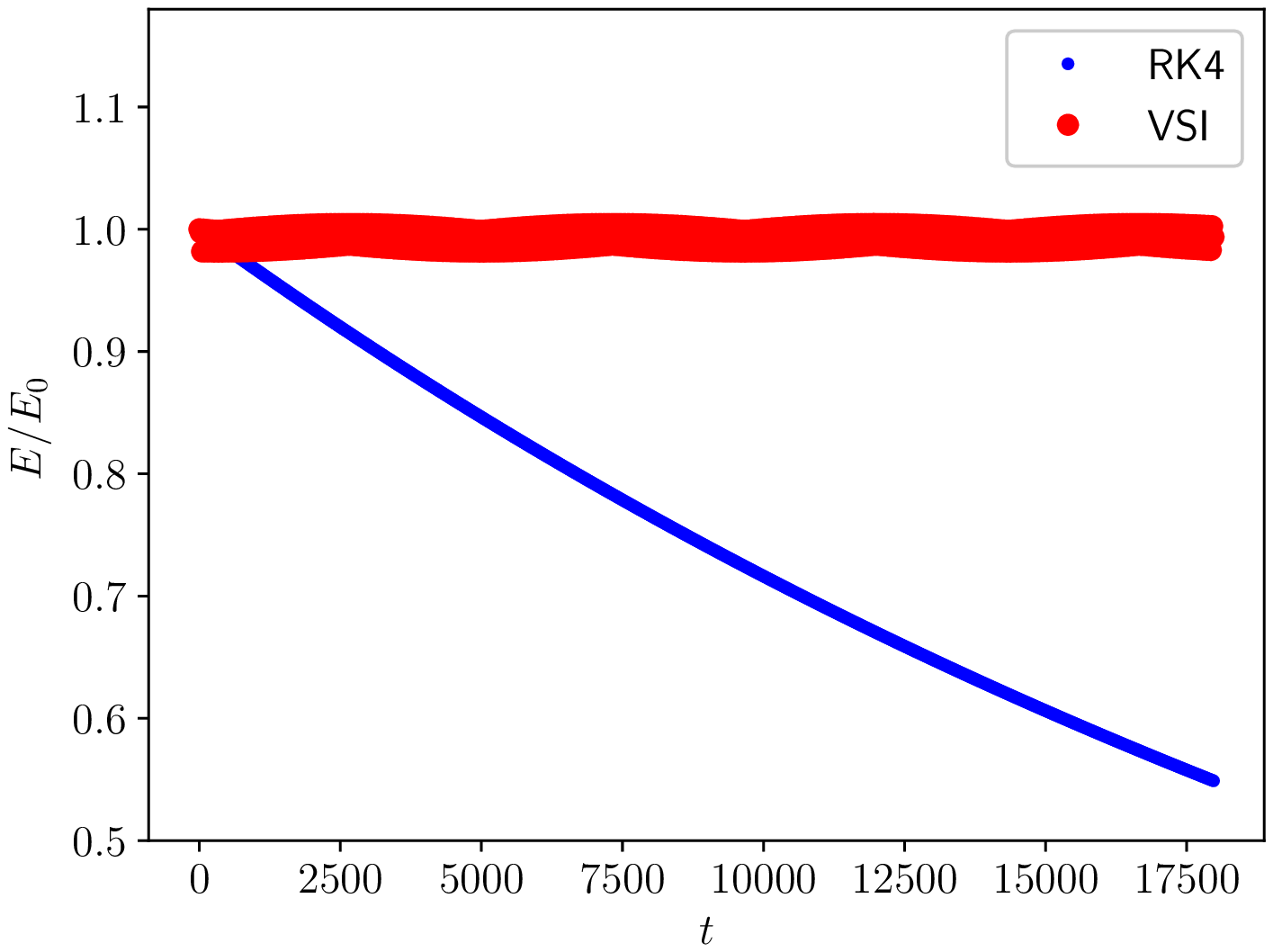}

}

\caption{The evolution of the harmonic oscillator in the $q-\dot{q}$ phase
space (a), and the evolution of total energy (b) solved by the RK4
and the VSI methods.}

\label{FigVSIRK} 
\end{figure}

To test the algorithm, we set $k=m=1$, $\Delta t=0.3$, $q_{0}=0$,
$\dot{q}_{1/2}=0.9$, $q_{1}=q_{0}+\dot{q}_{1/2}\Delta t=0.027$,
and iterate 60000 time steps. For comparison, we also used the standard
4th-order Runge-Kutta method (RK4) to solve the corresponding ODE.
The numerical results are shown in \FIG{FigVSIRK}. It is clear
that even for this simplest Hamiltonian system, the RK4 method failed
to preserve the orbit and energy for long-term dynamics. On the other
hand for the VSI method, the orbit is well preserved, and the error
of energy is bounded by a small value for all simulation time steps.

\section{Two Examples of Symplectic PIC methods}

In this section we introduce two simple ways to build symplectic PIC
integrators, which are based on variational and canonical symplectic
methods, respectively. Compared with the algorithms that will be introduced
in the next section, these two methods are not charge conservative,
and high-order schemes are generally implicit which are not convenient
for large scale simulations. But they are good examples to demonstrate
the general techniques to build symplectic PIC schemes.

According to discussions in the previous section, to construct a symplectic
PIC scheme, the starting point is the variational theory for the PIC
system. Consider a system that contains a collection of non-relativistic
charged particle and electromagnetic fields. The Lagrangian $L$ and
action integral $S$ are 
\begin{eqnarray}
L & = & \iiint\rmd\bfx\left(\frac{\epsilon_{0}}{2}\left(-\dot{\bfA}\left(\bfx\right)-\nabla\phi\left(\bfx\right)\right)^{2}-\frac{1}{2\mu_{0}}\left(\nabla\times\bfA\left(\bfx\right)\right)^{2}+\right.\nonumber \\
 &  & \left.\sum_{s,p}\delta\left(\bfx-\bfx_{sp}\right)\left(\frac{1}{2}m_{s}\dot{\bfx}_{sp}^{2}+q_{s}\bfA\left(\bfx\right)\cdot\dot{\bfx}_{sp}-q_{s}\phi\left(\bfx\right)\right)\right)~,\label{EqnLagVM}\\
S & = & \int_{0}^{T}L\rmd t~,
\end{eqnarray}
where $\epsilon_{0}$ and $\mu_{0}$ are the permittivity and permeability
in vacuum, $\bfA\left(\bfx\right)$ and $\phi\left(\bfx\right)$ are
the vector and scalar potentials of electromagnetic fields, $\bfx_{sp}$,
$m_{s}$ and $q_{s}$ are the location, mass and charge of the $p$-th
particle of species $s$, respectively. The equations of motion for
this particle-field system are determined by the variational principle,
\begin{eqnarray}
\frac{\delta S}{\delta\bfx_{sp}} & = & 0~,\label{EqnDELDX}\\
\frac{\delta S}{\delta\bfA} & = & 0~,\label{EqnDELDA}\\
\frac{\delta S}{\delta\phi} & = & 0~.\label{EqnDELDP}
\end{eqnarray}
Using the definition of electromagnetic fields $\bfE=-\dot{\bfA}-\nabla\phi$
and $\bfB=\nabla\times\bfA$, we can easily find that Eqs.\,\eqref{EqnDELDA}
and \eqref{EqnDELDP} are the Maxwell equations, and \EQ{EqnDELDX}
is Newton's equation for particles. Note that the dynamic equation
of the particle-field system Eqs. (\ref{EqnDELDX})-(\ref{EqnDELDP})
are gauge symmetric, i.e., they are invariant under the following
transform, 
\begin{eqnarray}
\bfA & \rightarrow & \bfA+\nabla\psi~,\\
\phi & \rightarrow & \phi-\frac{\partial\psi}{\partial t}~,
\end{eqnarray}
where $\psi(\bfx,t)$ is an arbitrary scalar function.

\subsection{Variational symplectic PIC algorithms}

Firstly, let us construct a PIC scheme based on the discrete variational
principle \cite{xiao2013variational}. This method does not make explicit
use of the Hamiltonian structure or any canonical transformation,
and all we need is to select a spatial discretization. It is the simplest
way to build a symplectic PIC scheme.

From the continuous Lagrangian given by \EQ{EqnLagVM}, we observe
that particle dynamics does not need any spatial discretization. Fields
and interaction between fields and particles can be discretized using
grids and interpolating functions. In the simplest case, a cubic mesh
can be used, and the fields are discretized on grid points as 
\begin{eqnarray}
\bfA_{i,j,k} & \approx & \bfA\left(i\Delta x,j\Delta x,k\Delta x\right)~,\\
\phi_{i,j,k} & \approx & \phi\left(i\Delta x,j\Delta x,k\Delta x\right)~,
\end{eqnarray}
where $i,j,k$ are grid indices in the $x,y,z$ directions and $0\leq i<N_{x}$,
$0\leq j<N_{y}$, $0\leq k<N_{z}$. Here, $N_{x},N_{y},N_{z}$ are
numbers of grids in the $x,y,z$ directions, respectively, and $\Delta x$
is the gird size. For discretizing the gradient and curl operators,
the forward difference is used. They are defined in Eqs.\,(\ref{EqnDEFGRADD})
and (\ref{EqnDEFcurld}). Since the locations of particles are not
always exactly on grid points, to calculate the interaction between
particles and electromagnetic fields an interpolating technique is
needed. The field interpolated at a given location $\bfx$ can be
written as 
\begin{eqnarray}
\bfA\left(\bfx\right) & \approx & \sum_{i,j,k}\bfA_{i,j,k}W\left(\bfx-\bfx_{i,j,k}\right)~,\label{EqnDIS2CONTA}\\
\phi\left(\bfx\right) & \approx & \sum_{i,j,k}\phi_{i,j,k}W\left(\bfx-\bfx_{i,j,k}\right)~,\label{EqnDIS2CONTP}
\end{eqnarray}
where $W$ is the interpolating function and $\bfx_{i,j,k}$ is the
location of the $i,j,k$-th grid point, i.e., $\bfx_{i,j,k}=\left(i\Delta x,j\Delta x,k\Delta x\right)$.
In general, $W$ should have following properties: 
\begin{enumerate}
\item It is normalized, i.e., for any $\bfx$, $\sum_{i,j,k}W\left(\bfx_{i,j,k}+\bfx\right)=1$; 
\item It is symmetric, i.e., for any $\bfx$, $W\left(\bfx\right)=W\left(-\bfx\right)$; 
\item It is localized, i.e., when $|\bfx|>a$, $W\left(\bfx\right)=0$,
where $a$ is the size of the interpolating function; 
\item It is smooth for reducing the numerical noise and improving the accuracy. 
\end{enumerate}
As an example, the following $W$ given by Ref.~\cite{xiao2013variational}
is adopted here, 
\begin{eqnarray}
W\left(\bfx\right) & = & W_{1}\left(x\right)W_{1}\left(y\right)W_{1}\left(z\right)~,\\
W_{1}\left(x\right) & =\nonumber 
\end{eqnarray}
\begin{eqnarray}
\left\{ \begin{array}{lc}
0~, & x>2~,\\
\frac{15}{1024}x^{8}-\frac{15}{128}x^{7}+\frac{49}{128}x^{6}-\frac{21}{32}x^{5}+\frac{35}{64}x^{4}-x+1~, & 1<x\leq2~,\\
-\frac{15}{1024}x^{8}-\frac{15}{128}x^{7}+\frac{7}{16}x^{6}-\frac{21}{32}x^{5}+\frac{175}{256}x^{4}-\frac{105}{128}x^{2}+\frac{337}{512}~, & 0<x\leq1~,\\
-\frac{15}{1024}x^{8}+\frac{15}{128}x^{7}+\frac{7}{16}x^{6}+\frac{21}{32}x^{5}+\frac{175}{256}x^{4}-\frac{105}{128}x^{2}+\frac{337}{512}~, & -1<x\leq0~,\\
\frac{15}{1024}x^{8}+\frac{15}{128}x^{7}+\frac{49}{128}x^{6}+\frac{21}{32}x^{5}+\frac{35}{64}x^{4}+x+1~, & -2<x\leq-1~,\\
0~, & x\leq-2~.
\end{array}\right.\label{EqnW1}
\end{eqnarray}
This interpolating function is smooth to the third order and piecewise
polynomial. It satisfies the above four properties. More discussions
on interpolating functions can be found in Refs.~\cite{lucy1977numerical,monaghan1992smoothed,liu2003smoothed,price2004smoothed}.
To simplify the scheme, we select the temporal gauge $\phi=0$. In
this gauge the electromagnetic fields are 
\begin{eqnarray}
\bfE & = & -\frac{\partial\bfA}{\partial t}~,\label{EqnEeqDADT}\\
\bfB & = & \nabla\times\bfA~.\label{EqnBeqDA}
\end{eqnarray}
Using temporal gauge to investigate electrostatic problems may result
in numerical overflow due to the continuously increasing vector potential.
However in most situations the plasma is quasi-neutral, which means
electrostatic field is small, and using temporal gauge is appropriate.
Now the discretized Lagrangian is 
\begin{eqnarray}
 &  & L_{d}\left(\bfx_{sp},\dot{\bfx}_{sp},\bfA_{i,j,k},\dot{\bfA}_{i,j,k}\right)=\sum_{i,j,k}\left(\frac{1}{2}\left(-\dot{\bfA}_{i,j,k}\right)^{2}-\frac{1}{2}\left(\CURLD\bfA_{i,j,k}\right)^{2}\right)+\nonumber \\
 &  & \sum_{s}\left(\frac{1}{2}m_{s}\dot{\bfx}_{sp}^{2}+q_{s}\sum_{i,j,k}\bfA_{i,j,k}W\left(\bfx_{i,j,k}-\bfx_{sp}\right)\cdot\dot{\bfx}_{sp}\right)~.\label{EqnLagDVM}
\end{eqnarray}
For the 1st-order variational integrator, the discretized action is
\begin{eqnarray}
S_{d} & = & \sum_{l=0}^{N-1}L_{d}\left(l,l+1\right)\Delta t~,
\end{eqnarray}
where 
\begin{eqnarray}
L_{d}\left(l,l+1\right) & = & L_{d}\left(\bfx_{sp,l},\frac{\bfx_{sp,l+1}-\bfx_{sp,l}}{\Delta t},\bfA_{i,j,k,l},\frac{\bfA_{i,j,k,l+1}-\bfA_{i,j,k,l}}{\Delta t}\right)~.
\end{eqnarray}
The final iteration scheme are given by the discrete Euler-Lagrange
equations, 
\begin{eqnarray}
\frac{\partial S_{d}}{\partial\bfx_{sp,l}} & = & 0~,\\
\frac{\partial S_{d}}{\partial\bfA_{i,j,k,l}} & = & 0~,
\end{eqnarray}
which are 
\begin{eqnarray}
m_{s}\frac{\bfx_{sp,l+1}-2\bfx_{sp,l}+\bfx_{sp,l-1}}{\Delta t^{2}} & = & q_{s}\frac{\left(\bfA_{l-1}\left(\bfx_{sp,l-1}\right)-\bfA_{l}\left(\bfx_{sp,l}\right)\right)}{\Delta t}+\label{EqnDELLORENTZ}\\
 &  & q_{s}\nabla\bfA_{l}\left(\bfx_{sp,l}\right)\cdot\frac{\bfx_{sp,l+1}-\bfx_{sp,l}}{\Delta t}~,\\
\frac{\bfA_{i,j,k,l+1}-2\bfA_{i,j,k,l}+\bfA_{i,j,k,l-1}}{\Delta t^{2}} & = & -\CURLDP\CURLD\bfA_{i,j,k,l}+\bfJ_{i,j,k,l}~,\label{EqnDSDELDA}
\end{eqnarray}
where 
\begin{eqnarray}
\bfA_{l}\left(\bfx\right) & = & \sum_{i,j,k}\bfA_{i,j,k,l}W\left(\bfx-\bfx_{i,j,k}\right)~,\\
\CURLDP\CURLD\bfA_{i,j,k,l} & = & \frac{\partial}{\partial\bfA_{i,j,k,l}}\sum_{i',j',k'}\frac{1}{2}\left(\CURLD\bfA\right)_{i',j',k',l}^{2}~,\\
\bfJ_{i,j,k,l} & = & \sum_{sp}q_{s}\frac{\bfx_{sp,l+1}-\bfx_{sp,l}}{\Delta t}W\left(\bfx_{i,j,k}-\bfx_{sp,l}\right)~.\label{EqnDELJ}
\end{eqnarray}
Equation (\ref{EqnDELLORENTZ}) is a $3\times3$ linear equation system
for $\bfx_{sp,l+1}$, and \EQ{EqnDSDELDA} is an explicit equation
of $\bfA_{i,j,k,l+1}$. The final algorithm is: 
\begin{enumerate}
\item $\bfA_{i,j,k,l-1}$, $\bfA_{i,j,k,l}$, $\bfx_{sp,l-1}$, $\bfx_{sp,l}$
are known for the $l$-th time step; 
\item Calculate $\bfx_{sp,l+1}$ using \EQ{EqnDELLORENTZ}; 
\item Calculate $\bfJ_{i,j,k,l}$ using \EQ{EqnDELJ}; 
\item Calculate $\bfA_{i,j,k,l+1}$ using \EQ{EqnDSDELDA}; 
\item Set $l=l+1$ and go to step 1. 
\end{enumerate}
Note that $W(\bfx)$ is only locally non-zero near $\bfx$, thus the
algorithm is locally explicit and convenient to compute in parallel.

\subsection{Canonical symplectic PIC algorithm}

Another way to build symplectic PIC algorithms is using the canonical
symplectic method \cite{qin2016canonical}. The starting point is
the discrete Lagrangian for the particle-field system under the temporal
gauge. According to the Legendre transformation Eq. (\ref{EqnLEGEND}),
the canonical momenta $(\bfp_{sp},\bfY_{i,j,k})$ are 
\begin{eqnarray}
\bfp_{sp} & = & \frac{\partial L_{d}}{\partial\dot{\bfx}_{sp}}~,\\
\bfY_{i,j,k} & = & \frac{\partial L_{d}}{\partial\dot{\bfA}_{i,j,k}}~.
\end{eqnarray}
The Hamiltonian in terms of the canonical pairs are given by \EQ{EqnHMT},
\begin{eqnarray}
H_{d}\left(\bfx_{sp},\bfp_{sp},\bfA_{i,j,k},\bfY_{i,j,k}\right) & = & \sum_{sp}\frac{1}{2}\frac{\left(\bfp_{sp}-q_{s}\bfA\left(\bfx_{sp}\right)\right)^{2}}{m_{s}}+\nonumber \\
 &  & \sum_{i,j,k}\frac{1}{2}\left(\frac{\bfY_{i,j,k}^{2}}{\Delta V}+\Delta V\left(\CURLD\bfA\right)_{i,j,k}^{2}\right)~,
\end{eqnarray}
where 
\begin{eqnarray}
\bfA\left(\bfx\right) & = & \sum_{i,j,k}\bfA_{i,j,k}W\left(\bfx-\bfx_{i,j,k}\right)~,
\end{eqnarray}
Here, the discrete $\CURLD$ operator is defined in \EQ{EqnDEFcurld}.
Using the 1st-order symplectic Euler method given by \EQ{EqnSEULER},
we obtain the 1st-order canonical symplectic PIC scheme, 
\begin{eqnarray}
\frac{\bfx_{sp,l+1}-\bfx_{sp,l}}{\Delta t} & = & \frac{\bfp_{sp,l+1}-\bfA_{l}\left(\bfx_{sp,l}\right)}{m_{s}}~,\\
\frac{\bfp_{sp,l+1}-\bfp_{sp,l}}{\Delta t} & = & q_{s}\nabla\bfA_{l}\left(\bfx_{sp,l}\right)\cdot\left(\bfp_{sp,l+1}-q_{s}\bfA_{l}\left(\bfx_{sp,l}\right)\right)~,\label{EqnBFP}\\
\frac{\bfA_{i,j,k,l+1}-\bfA_{i,j,k,l}}{\Delta t} & = & \frac{\bfY_{i,j,k,l+1}}{\Delta V}~,\\
\frac{\bfY_{i,j,k,l+1}-\bfY_{i,j,k,l}}{\Delta t} & = & \Delta V\left(\CURLDP\CURLD\bfA\right)_{i,j,k,l}+\nonumber \\
 &  & \sum_{sp}q_{s}W\left({\bfx_{sp,l}-\bfx_{i,j,k}}\right)\left(\bfp_{sp,l}-q_{s}\bfA_{l,l}\left(\bfx_{sp,l}\right)\right)~.
\end{eqnarray}
Here $\CURLDP$ is the transpose of $\CURLD$, as defined in Eq.\,\eqref{EqnDEFcurldp}.
We should note that Eq. \eqref{EqnBFP} is a $3\times3$ linear equation
array of $\bfp_{sp,l+1}$ which is easy to solve, so the whole scheme
is also explicit.

It can be proved that this 1st-order canonical symplectic PIC scheme
is equivalent to the 1st-order variational symplectic PIC scheme Eqs.~(\ref{EqnDELLORENTZ})
and (\ref{EqnDSDELDA}) under the following transformation, 
\begin{eqnarray}
\frac{\bfx_{sp,l+1}-\bfx_{sp,l}}{\Delta t} & = & \frac{\bfp_{sp,l+1}-\bfA_{l}\left(\bfx_{sp,l}\right)}{m_{s}}~,\\
\frac{\bfA_{i,j,k,l+1}-\bfA_{i,j,k,l}}{\Delta t} & = & \frac{\bfY_{i,j,k,l+1}}{\Delta V}~.
\end{eqnarray}

\section{Structure-Preserving Geometric PIC Algorithms}

The variational symplectic and the canonical symplectic PIC schemes
discussed above have good long-term conservative properties that conventional
PIC schemes do not have. However, they can further improved to preserve
more geometric structures and to increase accuracy. In this section,
we will describe explicit high-order non-canonical symplectic PIC
algorithms that admit a discrete gauge symmetry and a local charge
conservation law. To preserve geometric structures such as the symplectic
structure and the gauge symmetry, spatial discretization techniques
such as discrete exterior calculus (DEC) \cite{hirani2003discrete,stern2007geometric},
Whitney interpolating forms \cite{whitney1957geometric,hirani2003discrete,desbrun2008discrete},
and finite element exterior calculus (FEEC) \cite{he2016hamiltonian,kraus2017gempic}
can be used. Here, we will adopt the high-order Whitney interpolating
forms developed in Ref.\,\cite{xiao2015explicit}. The algorithm
is based on the discrete non-canonical Poisson bracket for the VM
system first given in Ref.\,\cite{xiao2015explicit}, which automatically
satisfies the Jacobi identity since it is derived from a discrete
variational principle. The explicit high-order non-canonical Hamiltonian
splitting method for the VM system discovered by He et al. \cite{he2015hamiltonian,he2017explicit}
is used. The connection between discrete gauge symmetry and discrete
charge conservation law is first pointed out by Squire et al. \cite{squire2012geometric}.
We will present a detailed proof of this fact with the help of the
Whitney interpolating forms. We will show that the geometric PIC algorithms
developed are gauge-symmetric and thus automatically satisfies the
charge conservation law. It is also shown that the 1st-order and the
2nd-order geometric PIC algorithms can be equivalently constructed
by the discrete variational approach. In addition, a relativistic
structure-preserving geometric PIC algorithm is given using the variational
approach. Two numerical examples are given to demonstrate the advantages
of the structure-preserving geometric PIC algorithms over the conventional
PIC schemes.

\subsection{Explicit high-order non-canonical symplectic PIC schemes}

\label{SecEHNCSPIC}

The starting point of the non-canonical PIC method is also the variational
principle of the particle-field system. The Lagrangian density of
the system is 
\begin{eqnarray}
L & = & \iiint\rmd\bfx\left(\frac{\epsilon_{0}}{2}\left(-\dot{\bfA}\left(\bfx\right)-\nabla\phi\left(\bfx\right)\right)^{2}-\frac{1}{2\mu_{0}}\left(\nabla\times\bfA\left(\bfx\right)\right)^{2}+\right.\nonumber \\
 &  & \left.\sum_{s}\delta\left(\bfx-\bfx_{sp}\right)\left(\frac{1}{2}m_{s}\dot{\bfx}_{sp}^{2}+q_{s}\bfA\left(\bfx\right)\cdot\dot{\bfx}_{sp}-q_{s}\phi\left(\bfx\right)\right)\right)~.
\end{eqnarray}
Fields are discretized over a cubic grid mesh, and a interpolating
technique is needed to calculate values of fields at the location
of particles. It is well-known that electromagnetic field has an intrinsic
differential form structure. Thus, we adopt a spatial discretization
scheme using the Whitney interpolating forms to preserve this geometric
structure. The resulting Lagrangian is 
\begin{eqnarray}
L_{sd}=\frac{1}{2}\left(\sum_{J}\left(-\dot{\bfA}_{J}-\sum_{I}{\GRADD}_{J,I}\phi_{I}\right)^{2}-\sum_{K}\left(\sum_{J}\CURLD_{K,J}\bfA_{J}\right)^{2}\right)\Delta V+\nonumber \\
\sum_{s}\left(\frac{1}{2}m_{s}\dot{\bfx}_{sp}^{2}+q_{s}\left(\dot{\bfx}_{sp}\cdot\sum_{J}\WONE{\bfx_{sp}}\bfA_{J}-\sum_{I}\WZERO{\bfx_{sp}}\phi_{I}\right)\right)~,
\end{eqnarray}
where $\CURLD$ and $\GRADD$ are discrete curl and gradient operators,
$I,J,K,L$ are the indices for the discrete 0, 1, 2, 3-forms, respectively,
$\WZERO{\bfx}$, $\WONE{\bfx}$ and $\WTWO{\bfx}$ are interpolating
functions for discrete 0-forms, 1-forms and 2-forms. To preserve the
geometric structure, interpolating functions and discrete operators
are required to satisfy following properties, 
\begin{eqnarray}
\nabla\sum_{I}\WZERO{\bfx}\phi_{I}=\sum_{I,J}\WONE{\bfx}{\GRADD}_{J,I}\phi_{I}~,\\
\nabla\times\sum_{J}\WONE{\bfx}\bfA_{J}=\sum_{J,K}\WTWO{\bfx}{\CURLD}_{K,J}\bfA_{J}~.\label{EqnD1to2FORM}
\end{eqnarray}
These relations are the same as the properties of Whitney interpolating
forms \cite{whitney1957geometric,hirani2003discrete,desbrun2008discrete,squire2012geometric},
in which the interpolating map $\phi_{W}$ provides a mechanism to
define continuous forms from the discrete ones. For the present application,
$\phi_{W}$ is either $\WZERO{\bfx}$ or $\WONE{\bfx}$. The map and
discrete exterior difference operators satisfy the following rule,
\begin{eqnarray}
\phi_{W}\bfd_{\rmd}\alpha=\bfd\phi_{W}\alpha~,
\end{eqnarray}
where $\bfd$ and $\bfd_{\rmd}$ are the continuous exterior differential
and the discrete exterior difference, respectively, and $\alpha$
is an arbitrary discrete form. The original Whitney interpolating
map is defined in simplex meshes (e.g., triangle meshes and tetrahedron
meshes) and can only affect one grid cell. We have built a new set
of interpolating maps which is suitable for cubic meshes and is able
to perform high-order interpolation over multiple grid cells \cite{xiao2015explicit}.
The resulting discrete difference operators and interpolating maps
are listed in Eqs.~(\ref{EqnDEFGRADD}-\ref{EqnDEFW3}). We should
emphasize that for forms with different orders (e.g. vector potential
and magnetic field), the resulting interpolating functions are different.
Even for different components of the same field in different directions,
the interpolating functions are also different. This is very different
from conventional interpolating techniques \cite{Dawson1983,birdsall1991plasma,hockney1988computer}.
As a result of the structure-preserving properties, the discrete exterior
difference operators satisfy 
\begin{eqnarray}
\sum_{J,I}\CURLD_{K,J}{\GRADD}_{J,I}\phi_{I} & = & 0~,\label{EqnCURLDGRADDZERO}\\
\sum_{K,J}\DIVD_{L,K}\CURLD_{K,J}\bfA_{J} & = & 0~,\label{EqnDIVDCURLDZERO}
\end{eqnarray}
for any discrete 0-form $\phi_{I}$ and 1-form $\bfA_{J}$. These
are obviously the discrete analogs of 
\begin{eqnarray}
\nabla\times\left(\nabla\phi\right) & = & 0~,\\
\nabla\cdot\left(\nabla\times\bfA\right) & = & 0~.
\end{eqnarray}

The corresponding equations of motion are 
\begin{eqnarray}
\frac{\delta S_{sd}}{\delta\bfA_{J}} & = & 0~,\label{EqnMaxwellEqn}\\
\frac{\delta S_{sd}}{\delta\phi_{I}} & = & 0~,\label{EqnPoisson}\\
\frac{\delta S_{sd}}{\delta\bfx_{sp}} & = & 0~,\label{EqnLorentz}
\end{eqnarray}
or more specifically, 
\begin{eqnarray}
\ddot{\bfx}_{sp} & = & \frac{q_{s}}{m_{s}}\left(\dot{\bfx}_{sp}\times\left(\nabla\times\sum_{J}\bfA_{J}\WONE{\bfx_{sp}}\right)-\right.\nonumber \\
 &  & \left.\sum_{J}\dot{\bfA}_{J}\WONE{\bfx_{sp}}-\sum_{I}\nabla\phi_{I}\WZERO{\bfx_{sp}}\right)~,\label{EqnDISLORENTZ}\\
\ddot{\bfA}_{J}+\sum_{I}{\GRADD}_{J,I}\dot{\phi}_{I} & = & -\CURLDP\CURLD\bfA_{J}+\sum_{sp}q_{s}\dot{\bfx}_{sp}\WONE{\bfx_{sp}}~,\\
0 & = & \sum_{J}{\GRADD}_{J,I}\left(\dot{\bfA}_{J}+\sum_{I}{\GRADD}_{J,I}\phi_{I}\right)-\sum_{sp}q_{s}\WZERO{\bfx_{sp}},\label{EqnDISGAUSS}
\end{eqnarray}
where $S_{sd}$ is the spatially discretized action integral, 
\begin{eqnarray}
S_{sd}=\int\rmd tL_{sd}~.
\end{eqnarray}
Using the property of interpolating maps, we can find that the equations
of motion are gauge-free, or symmetric with respect to the electromagnetic
gauge, which means that under the following transformation of vector
and scalar potentials defined on the grid,
\begin{eqnarray}
\bfA_{J} & \rightarrow & \bfA_{J}+\sum_{I}{\GRADD}_{J,I}\psi_{I}~,\label{eq:gs1}\\
\phi_{I} & \rightarrow & \phi_{I}-\frac{\partial\psi_{I}}{\partial t}~,\label{eq:gs2}
\end{eqnarray}
where $\psi_{I}$ is an arbitrary discrete 0-form, Eqs.~(\ref{EqnDISLORENTZ}-\ref{EqnDISGAUSS})
do not change.

To obtain an algorithm, we follow Ref.\,\cite{xiao2015explicit}
to derive an discrete non-canonical Poisson bracket. Because the dynamical
equations are gauge-free, we can choose the temporal gauge $\phi_{I}=0$
for simplicity. The corresponding Lagrangian 1-form \cite{marsden2013introduction}
is 
\begin{eqnarray}
\gamma=\frac{\partial L_{sd}}{\partial\dot{q}}\bfd q-H_{sd}\bfd t
\end{eqnarray}
where $q=[\bfA_{J},\bfx_{sp}]$, and $H_{sd}$ is spatially discretized
Hamiltonian determined by the Legendre transform, 
\begin{eqnarray}
H_{sd} & = & \frac{\partial L_{sd}}{\partial\dot{q}}\dot{q}-L_{sd}\nonumber \\
 & = & \frac{1}{2}\Delta V\left(\sum_{J}\dot{\bfA}_{J}^{2}+\sum_{K}\left(\sum_{J}\CURLD_{K,J}\bfA_{J}\right)^{2}\right)+\sum_{s}\frac{1}{2}m_{s}\dot{\bfx}_{sp}^{2}~.
\end{eqnarray}
The symplectic 2-form of this system is 
\begin{eqnarray}
\Omega=\bfd\left(\frac{\partial L_{sd}}{\partial\dot{q}}\bfd q\right)~, & \text{}\label{eq:Omega}
\end{eqnarray}
from which the non-canonical Poisson bracket is 
\begin{eqnarray}
\left\{ F,G\right\} =\left[\frac{\partial F}{\partial\bfA_{J}},\frac{\partial F}{\partial\bfx_{sp}},\frac{\partial F}{\partial\dot{\bfA}_{J}},\frac{\partial F}{\partial\dot{\bfx}_{sp}}\right]\Omega^{-1}\left[\frac{\partial G}{\partial\bfA_{J}},\frac{\partial G}{\partial\bfx_{sp}},\frac{\partial G}{\partial\dot{\bfA}_{J}},\frac{\partial G}{\partial\dot{\bfx}_{sp}}\right]^{T},
\end{eqnarray}
or more specifically 
\begin{eqnarray}
\left\{ F,G\right\}  & = & \frac{1}{\Delta V}\sum_{J}\left(\frac{\partial F}{\partial\bfA_{J}}\cdot\frac{\partial G}{\partial\dot{\bfA}_{J}}-\frac{\partial F}{\partial\dot{\bfA}_{J}}\cdot\frac{\partial G}{\partial\bfA_{J}}\right)+\nonumber \\
 &  & \sum_{s}\frac{1}{m_{s}}\left(\frac{\partial F}{\partial\bfx_{sp}}\cdot\frac{\partial G}{\partial\dot{\bfx}_{sp}}-\frac{\partial F}{\partial\dot{\bfx}_{sp}}\cdot\frac{\partial G}{\partial\bfx_{sp}}\right)+\nonumber \\
 &  & \sum_{s}\frac{q_{s}}{m_{s}\Delta V}\left(\frac{\partial G}{\partial\dot{\bfx}_{sp}}\cdot\sum_{J}W_{\sigma_{1J}}\left(\bfx_{sp}\right)\frac{\partial F}{\partial\dot{\bfA}_{J}}-\frac{\partial F}{\partial\dot{\bfx}_{sp}}\cdot\sum_{J}W_{\sigma_{1J}}\left(\bfx_{sp}\right)\frac{\partial G}{\partial\dot{\bfA}_{J}}\right)+\nonumber \\
 &  & -\sum_{s}\sum_{J}\frac{q_{s}}{m_{s}^{2}}\frac{\partial F}{\partial\dot{\bfx}_{sp}}\cdot\left(\left(\nabla\times W_{\sigma_{1J}}\left(\bfx_{sp}\right)\bfA_{J}\right)\times\frac{\partial G}{\partial\dot{\bfx}_{sp}}\right)~.
\end{eqnarray}
Now we introduce discrete electromagnetic fields $\bfE_{J}$ and $\bfB_{K}$
as follows, 
\begin{eqnarray}
\bfE_{J} & = & -\dot{\bfA}_{J}~,\\
\bfB_{K} & = & \sum_{J}\CURLD_{K,J}\bfA_{J}~.
\end{eqnarray}
Making use of the transformation of partial derivatives with respect
to $\bfA_{J},\dot{\bfA}_{J}$ and $\bfE_{J},\bfB_{K}$, 
\begin{eqnarray}
\frac{\partial F}{\partial\bfA_{J}} & = & \sum_{K}\frac{\partial F}{\partial\bfB_{K}}\CURLD_{K,J}~,\label{eq:20}\\
\frac{\partial F}{\partial\dot{\bfA}_{J}} & = & -\frac{\partial F}{\partial\bfE_{J}}~,
\end{eqnarray}
as well as properties of Whitney interpolating maps, we can express
the Poisson bracket in terms of $[\bfE_{J},\bfB_{K},\bfx_{sp},\dot{\bfx}_{sp}]$:
\begin{eqnarray}
\left\{ F,G\right\}  & = & \frac{1}{\Delta V}\sum_{J}\left(\frac{\partial F}{\partial\bfE_{J}}\cdot\sum_{K}\frac{\partial G}{\partial\bfB_{K}}\CURLD_{K,J}-\sum_{K}\frac{\partial F}{\partial\bfB_{K}}\CURLD_{K,J}\cdot\frac{\partial G}{\partial\bfE_{J}}\right)+\nonumber \\
 &  & \sum_{s}\frac{1}{m_{s}}\left(\frac{\partial F}{\partial\bfx_{sp}}\cdot\frac{\partial G}{\partial\dot{\bfx}_{sp}}-\frac{\partial F}{\partial\dot{\bfx}_{sp}}\cdot\frac{\partial G}{\partial\bfx_{sp}}\right)+\nonumber \\
 &  & \sum_{s}\frac{q_{s}}{m_{s}\Delta V}\left(\frac{\partial F}{\partial\dot{\bfx}_{sp}}\cdot\sum_{J}W_{\sigma_{1J}}\left(\bfx_{sp}\right)\frac{\partial G}{\partial\bfE_{J}}-\frac{\partial G}{\partial\dot{\bfx}_{sp}}\cdot\sum_{J}W_{\sigma_{1J}}\left(\bfx_{sp}\right)\frac{\partial F}{\partial\bfE_{J}}\right)+\nonumber \\
 &  & -\sum_{s}\frac{q_{s}}{m_{s}^{2}}\frac{\partial F}{\partial\dot{\bfx}_{sp}}\cdot\left[\sum_{K}W_{\sigma_{2K}}\left(\bfx_{sp}\right)\bfB_{K}\right]\times\frac{\partial G}{\partial\dot{\bfx}_{sp}}~.\label{EqnPoissonBraketEB}
\end{eqnarray}
And the corresponding Hamiltonian is 
\begin{eqnarray}
H_{sd}=\frac{1}{2}\left(\Delta V\sum_{J}\bfE_{J}^{2}+\Delta V\sum_{K}\bfB_{K}^{2}+\sum_{sp}m_{s}\dot{\bfx}_{sp}^{2}\right)~.\label{EqnHamitEB}
\end{eqnarray}
The dynamics are then given by the Hamiltonian equations, 
\begin{eqnarray}
\dot{F}=\{F,H_{sd}\}~,
\end{eqnarray}
where 
\begin{eqnarray}
F=[\bfE_{J},\bfB_{K},\bfx_{sp},\dot{\bfx}_{sp}]~.
\end{eqnarray}
Note that the non-canonical Poisson bracket is defined by the 2-form
$\Omega$ given by Eq.\,\eqref{eq:Omega} and thus automatically
satisfies the Jacobi identity. This finite dimensional non-canonical
Hamiltonian system can be viewed as a Hamiltonian discretization of
the continuous VM system \cite{morrison1980maxwell,marsden1982hamiltonian,morrison1998hamiltonian}.

Because the Hamiltonian dose not explicitly depend on time $t$, the
total energy is dynamic invariant. Interestingly, there is also a
discrete local energy conservation law for this discrete non-canonical
Hamiltonian system. Detail discussions of the discrete local energy
conservation law can be found in Ref.~\cite{xiao2017local}.

Generally speaking, structure-preserving algorithms for non-canonical
Hamiltonian systems are difficult to construct. However we found \cite{xiao2015explicit}
that the high-order explicit Hamiltonian splitting method discovered
by He. et al~\cite{he2015hamiltonian} for the continuous VM system
is applicable to our discrete Hamiltonian system. In this method,
we split the Hamiltonian into five parts, 
\begin{eqnarray}
H_{sd}=H_{E}+H_{B}+H_{x}+H_{y}+H_{z}~,\\
H_{E}=\frac{1}{2}\Delta V\sum_{J}\bfE_{J}^{2}~,\\
H_{B}=\frac{1}{2}\Delta V\sum_{K}\bfB_{K}^{2}~,\\
H_{r}=\frac{1}{2}\sum_{sp}m_{s}\dot{r}_{sp}^{2},\quad\textrm{for }r\textrm{ in}~x,~y,~z~.
\end{eqnarray}
It turns out that all of the five Hamiltonian subsystems can be solved
analytically, and high-order algorithms for the entire system can
be built by compositions.

For $H_{E}$ and $H_{B}$, the corresponding dynamical equations are
$\dot{F}=\{F,H_{E}\}$ and $\dot{F}=\{F,H_{B}\}$, which are

\begin{equation}
\left\{ \begin{array}{ccl}
\dot{\bfE}_{J} & = & 0~,\\
\dot{\bfB}_{K} & = & -\sum_{J}\CURLD_{K,J}\bfE_{J}~,\\
\dot{\bfx}_{sp} & = & 0~,\\
\ddot{\bfx}_{sp} & = & \frac{q_{s}}{m_{s}}\sum_{J}W_{\sigma_{1J}}\left(\bfx_{sp}\right)\bfE_{J}~.
\end{array}\right.
\end{equation}
and

\begin{equation}
\left\{ \begin{array}{ccl}
\dot{\bfE}_{J} & = & \sum_{K}\CURLD_{K,J}\bfB_{K}~,\\
\dot{\bfB}_{K} & = & 0~,\\
\dot{\bfx}_{sp} & = & 0~,\\
\ddot{\bfx}_{sp} & = & 0~.
\end{array}\right.
\end{equation}
Their solution maps $\Theta_{E}\left(\Delta t\right)$ and $\Theta_{B}\left(\Delta t\right)$
are

\begin{equation}
\Theta_{E}\left(\Delta t\right):\left\{ \begin{array}{ccl}
\bfE_{J}\left(t+\Delta t\right) & = & \bfE_{J}\left(t\right)~,\\
\bfB_{K}\left(t+\Delta t\right) & = & \bfB_{K}\left(t\right)-\Delta t\sum_{J}\CURLD_{K,J}\bfE_{J}(t)~,\\
\bfx_{sp}\left(t+\Delta t\right) & = & \bfx_{sp}\left(t\right)~,\\
\dot{\bfx}_{sp}\left(t+\Delta t\right) & = & \dot{\bfx}_{sp}\left(t\right)+\frac{q_{s}}{m_{s}}\Delta t\sum_{J}W_{\sigma_{1J}}\left(\bfx_{sp}(t)\right)\bfE_{J}(t)~,
\end{array}\right.
\end{equation}
and 
\begin{equation}
\Theta_{B}\left(\Delta t\right):\left\{ \begin{array}{ccl}
\bfE_{J}\left(t+\Delta t\right) & = & \bfE_{J}\left(t\right)+\Delta t\sum_{K}\CURLD_{K,J}\bfB_{K}(t)~,\\
\bfB_{K}\left(t+\Delta t\right) & = & \bfB_{K}\left(t\right)~,\\
\bfx_{sp}\left(t+\Delta t\right) & = & \bfx_{sp}\left(t\right)~,\\
\dot{\bfx}_{sp}\left(t+\Delta t\right) & = & \dot{\bfx}_{sp}\left(t\right)~.
\end{array}\right.
\end{equation}
For $H_{x}$, the dynamical equation is $\dot{F}=\left\{ F,H_{x}\right\} $,
i.e., 
\begin{equation}
\left\{ \begin{array}{ccl}
\dot{\bfE}_{J} & = & -\sum_{s}\frac{q_{s}}{\Delta V}\dot{x}_{sp}\bfe_{x}W_{\sigma_{1J}}\left(\bfx_{sp}\right)~,\\
\dot{\bfB}_{K} & = & 0~,\\
\dot{\bfx}_{sp} & = & \dot{x}_{sp}\bfe_{x}~,\\
\ddot{\bfx}_{sp} & = & \frac{q_{s}}{m_{s}}\dot{x}_{sp}\bfe_{x}\times\sum_{K}W_{\sigma_{2K}}\left(\bfx_{sp}\right)\bfB_{K}~.
\end{array}\right.
\end{equation}
And the solution map $\Theta_{x}\left(\Delta t\right)$ is

\[
\Theta_{x}\left(\Delta t\right):\left\{ \begin{array}{ccl}
\bfE_{J}\left(t+\Delta t\right) & = & \bfE_{J}\left(t\right)-\\
 &  & \int_{0}^{\Delta t}dt'\sum_{s}\frac{q_{s}}{\Delta V}\dot{x}_{sp}(t)\bfe_{x}W_{\sigma_{1J}}\left(\bfx_{sp}\left(t\right)+\dot{x}_{sp}(t)t'\bfe_{x}\right)~,\\
\bfB_{K}\left(t+\Delta t\right) & = & \bfB_{K}\left(t\right)~,\\
\bfx_{sp}\left(t+\Delta t\right) & = & \bfx_{sp}\left(t\right)+\Delta t\dot{x}_{sp}(t)\bfe_{x}~,\\
\dot{\bfx}_{sp}\left(t+\Delta t\right) & = & \dot{\bfx}_{sp}\left(t\right)+\frac{q_{s}}{m_{s}}\dot{x}_{sp}(t)\bfe_{x}\times\\
 &  & \int_{0}^{\Delta t}dt'\sum_{K}W_{\sigma_{2K}}\left(\bfx_{sp}\left(t\right)+\dot{x}_{sp}(t)t'\bfe_{x}\right)\bfB_{K}(t)~.
\end{array}\right.
\]
Solution maps $\Theta_{y}$ and $\Theta_{z}$ of subsystems with Hamiltonian
$H_{y}$ and $H_{z}$ respectively can be obtained similarly (see
Eqs. (\ref{EQNTHETAY}, \ref{EQNTHETAZ})). It is clear that all these
solutions are symplectic-structure-preserving, and so are their compositions.
Therefore, we can build symplectic schemes for the entire system using
compositions. For example, a 1st-order scheme can be chosen as 
\begin{eqnarray}
\Theta_{1}\left(\Delta t\right)=\Theta_{x}\left(\Delta t\right)\Theta_{y}\left(\Delta t\right)\Theta_{z}\left(\Delta t\right)\Theta_{E}\left(\Delta t\right)\Theta_{B}\left(\Delta t\right)~,\label{EqnFSTHS}
\end{eqnarray}
a symmetric 2nd-order scheme can be constructed as 
\begin{eqnarray}
\Theta_{2}\left(\Delta t\right) & = & \Theta_{E}\left(\Delta t/2\right)\Theta_{x}\left(\Delta t/2\right)\Theta_{y}\left(\Delta t/2\right)\Theta_{z}\left(\Delta t/2\right)\Theta_{B}\left(\Delta t\right)\nonumber \\
 &  & \Theta_{z}\left(\Delta t/2\right)\Theta_{y}\left(\Delta t/2\right)\Theta_{x}\left(\Delta t/2\right)\Theta_{E}\left(\Delta t/2\right)~,\label{EqnHAMS2}
\end{eqnarray}
and a $2(l+1)$-th order scheme can be constructed from a $2l$-th
order scheme as \cite{yoshida1990construction} 
\begin{eqnarray}
\Theta_{2(l+1)}(\Delta t) & = & \Theta_{2l}(\alpha_{l}\Delta t)\Theta_{2l}(\beta_{l}\Delta t)\Theta_{2l}(\alpha_{l}\Delta t)~,\\
\alpha_{l} & = & 1/(2-2^{1/(2l+1)})~,\\
\beta_{l} & = & 1-2\alpha_{l}~.
\end{eqnarray}

Similar or identical structure-preserving geometric PIC algorithms
can also be constructed using the discrete variational method. To
achieve the charge-conservation property, a gauge-symmetric discrete
Lagrangian should be used. The key is to choose a proper integrating
path of charged particles. Squire et al. have found a polyline path
to achieve this goal \cite{squire2012geometric}. However, the scheme
is implicit. Here, we propose a zigzag path that will render an explicit
algorithm. The discrete action is 
\begin{equation}
S_{d}=\sum_{l=0}^{N_{t}-1}\Delta tL_{\textrm{dxzig}}\left(\bfx_{sp,l},\bfx_{sp,l+1},\bfA_{J,l},\bfA_{J,l+1},\phi_{I,l};\Delta t\right)~,\label{EqnSACT}
\end{equation}
\begin{eqnarray}
 &  & L_{\textrm{dxzig}}\left(\bfx_{sp,l},\bfx_{sp,l+1},\bfA_{J,l},\bfA_{J,l+1},\phi_{I,l};\Delta t\right)=L_{\textrm{dfield}}+\nonumber \\
 &  & \sum_{sp}\left(\frac{1}{2}m_{s}\left(\DDELTATA{\bfx_{sp,l+1}}\right)^{2}+q_{s}\left(\DDELTATA{{\bfx}_{sp,l+1}}\cdot\int_{0}^{1}\rmd\tau\bfA_{l}\left(\bfxzig\left(\bfx_{sp,l},\bfx_{sp,l+1},\tau\right)\right)\right)-\right.\nonumber \\
 &  & \left.q_{s}\sum_{I}\WZERO{\bfx_{sp,l}}\phi_{I,l}\right)~,\label{EqnSZIG}
\end{eqnarray}
\begin{equation}
L_{\textrm{dfield}}=\frac{1}{2}\Delta V\left(\sum\left(-\DDELTATA{\bfA_{J,l+1}}-\sum_{I}{\GRADD}_{J,I}\phi_{I,l+1}\right)^{2}-\sum_{K}\left(\sum_{J}\CURLD_{K,J}\bfA_{J,l}\right)^{2}\right)~,
\end{equation}
where $N_{t}$ is the number of time steps, and 
\begin{eqnarray}
\DDELTAT{f_{l}} & = & \frac{f_{l+1}-f_{l}}{\Delta t}~,\\
\DDELTATA{f_{l}} & = & \frac{f_{l}-f_{l-1}}{\Delta t}~,\\
\bfA_{l}\left(\bfx\right) & = & \left[\begin{array}{c}
A_{x,l}\left(\bfx\right),\\
A_{y,l}\left(\bfx\right),\\
A_{z,l}\left(\bfx\right)
\end{array}\right]=\sum_{J}\WONE{\bfx}\bfA_{J,l}~,\\
\bfxzig\left(\bfx_{1},\bfx_{2},\tau\right) & = & \left[\begin{array}{c}
\xzig\left(\bfx_{1},\bfx_{2},\tau\right),\\
\yzig\left(\bfx_{1},\bfx_{2},\tau\right),\\
\zzig\left(\bfx_{1},\bfx_{2},\tau\right)
\end{array}\right]=\left[\begin{array}{c}
\left(x_{1}+\tau\left(x_{2}-x_{1}\right),y_{1},z_{1}\right),\\
\left(x_{2},y_{1}+\tau\left(y_{2}-y_{1}\right),z_{1}\right),\\
\left(x_{2},y_{2},z_{1}+\tau\left(z_{2}-z_{1}\right)\right)
\end{array}\right]~,\\
\bfA_{l}\left(\bfxzig\left(\bfx_{1},\bfx_{2},\tau\right)\right) & = & \left[\begin{array}{c}
A_{x,l}\left(\xzig\left(\bfx_{1},\bfx_{2},\tau\right)\right),\\
A_{y,l}\left(\yzig\left(\bfx_{1},\bfx_{2},\tau\right)\right),\\
A_{z,l}\left(\zzig\left(\bfx_{1},\bfx_{2},\tau\right)\right)
\end{array}\right]~.
\end{eqnarray}
Discrete dynamical equations for the fields is obtained from the variation
of the discrete action $S_{d}$ with respect to $\bfA_{J,l}$, $\phi_{I,l}$,
which yields 
\begin{eqnarray}
\DDELTAT{-\DDELTATA{\bfA_{J,l}}-\sum_{I}{\GRADD}_{J,I}\phi_{I,l}}+\CURLD^{T}\CURLD\bfA_{J,l} & = & \bfJ_{J,l}~,\quad\label{EqnDEDTREL}\\
\sum_{J}{\GRADD}_{J,I}\left(-\DDELTATA{\bfA_{J,l}}+\sum_{I'}{\GRADD}_{J,I'}\phi_{I',l}\right) & = & \rho_{I,l}~,\label{EqnDERHOREL}
\end{eqnarray}
where 
\begin{eqnarray}
\rho_{I,l} & = & \sum_{sp}q_{s}\WZERO{\bfx_{sp,l}}~,\label{EqnRHODEF}\\
\bfJ_{J,l} & = & \sum_{sp}q_{s}\DDELTATA{\bfx_{sp,l+1}}\cdot\int_{0}^{1}\rmd\tau\WONE{\bfxzig\left(\bfx_{sp,l},\bfx_{sp,l+1},\tau\right)}~.\label{EqnCURRDEF}
\end{eqnarray}
We can also introduce discrete electromagnetic fields as 
\begin{eqnarray}
\bfE_{J,l} & = & -\DDELTATA{\bfA_{J,l}}-\sum_{I}\GRADD_{J,I}\phi_{I,l}~,\\
\bfB_{K,l} & = & \sum_{J}\CURLD_{K,J}\bfA_{J,l}~.
\end{eqnarray}
Then Eqs.~(\ref{EqnDEDTREL}) and (\ref{EqnDERHOREL}) become 
\begin{eqnarray}
\DDELTATA{\bfB_{K,l}} & = & -\sum_{J}\CURLD_{K,J}\bfE_{J,l}~,\label{EqnB}\\
\DDELTAT{\bfE_{J,l}} & = & \sum_{K}\CURLDP_{J,K}\bfB_{K,l}-\bfJ_{J,l}~,\label{EqnE}\\
\sum_{J}-{\GRADD}_{J,I}\bfE_{J,l} & = & \rho_{I,l}~.\label{EqnRHO}
\end{eqnarray}
In practice, the discrete Poisson equation, i.e., \EQ{EqnRHO},
is automatically satisfied when solving the discrete Maxwell equation,
due to the discrete charge conservation property, which will be discussed
later. This situation is similar to the continuous case, where the
Poisson equation is taken as an initial condition. However, we emphasize
that only when the PIC algorithms are able to preserve the discrete
charge conservation law, the discrete Poisson equation can be automatically
satisfied. Particles' equation of motion is obtained from variation
of the discrete $S_{d}$ with respect to $\bfx_{sp,l}$. For example,
the variation with respect to $x_{sp,l}$ gives the equation to determine
$x_{sp,l+1}$ , 
\begin{eqnarray}
 &  & m_{s}\frac{x_{sp,l+1}-2x_{sp,l}+x_{sp,l-1}}{q_{s}\Delta t^{2}}=\frac{1}{\Delta t}\int_{0}^{1}\rmd\tau A_{x,l-1}\left(\xzig\zigspmvar\right)\nonumber \\
 &  & -\frac{1}{\Delta t}\int_{0}^{1}\rmd\tau A_{x,l}\left(\xzig\zigspmvar\right)+\nonumber \\
 &  & \DDELTATA{\bfx_{sp,l}}\cdot\int_{0}^{1}\rmd\tau\left[\begin{array}{c}
\tau\partial_{x}A_{x,l-1}\left(\xzig\zigspmvar\right),\\
\partial_{x}A_{y,l-1}\left(\yzig\zigspmvar\right),\\
\partial_{x}A_{z,l-1}\left(\zzig\zigspmvar\right)
\end{array}\right]+\nonumber \\
 &  & \frac{x_{sp,l}-x_{sp,l-1}}{\Delta t}\int_{0}^{1}\rmd\tau\left(1-\tau\right)\partial_{x}{A_{x}}_{l}\left(\zzig\zigspvar\right)-\nabla\sum_{I}W_{\sigma_{0I}}\left(\bfx_{sp,l}\right)\phi_{I,l}~.\label{EqnDSREDX}
\end{eqnarray}
Using following identities, 
\begin{eqnarray}
 &  & \frac{\rmd}{\rmd\tau}\left(\frac{\tau}{\Delta t}A_{x,l-1}\left(\xzig\zigspmvar\right)\right)\nonumber \\
= &  & \frac{x_{sp,l}-x_{sp,l-1}}{\Delta t}\tau\partial_{x}A_{x,l-1}\left(\xzig\zigspmvar\right)+\frac{1}{\Delta t}A_{x,l-1}\left(\xzig\zigspmvar\right)~,\\
 &  & \frac{\rmd}{\rmd\tau}\left(\frac{1-\tau}{\Delta t}A_{x,l}\left(\xzig\zigspvar\right)\right)\nonumber \\
= &  & \frac{x_{sp,l+1}-x_{sp,l}}{\Delta t}\left(1-\tau\right)\partial_{x}A_{x,l}\left(\xzig\zigspvar\right)-\frac{1}{\Delta t}A_{x,l}\left(\xzig\zigspvar\right)~,\\
 &  & \frac{\rmd}{\rmd\tau}A_{x,l-1}\left(\yzig\zigspmvar\right)=\frac{y_{sp,l}-y_{sp,l-1}}{\Delta t}\partial_{y}A_{x,l-1}\left(\yzig\zigspmvar\right)~,\\
 &  & \frac{\rmd}{\rmd\tau}A_{x,l-1}\left(\zzig\zigspmvar\right)=\frac{z_{sp,l}-z_{sp,l-1}}{\Delta t}\partial_{z}A_{x,l-1}\left(\zzig\zigspmvar\right)~,
\end{eqnarray}
we can simplify \EQ{EqnDSREDX} as 
\begin{eqnarray}
 &  & m_{s}\frac{x_{sp,l+1}-2x_{sp,l}+x_{sp,l-1}}{q_{s}\Delta t^{2}}=\left.\left(\frac{\tau}{\Delta t}A_{x,l-1}\left(\xzig\zigspmvar\right)\right)\right|_{\tau=0}^{\tau=1}\nonumber \\
 &  & +\left.\left(\frac{1-\tau}{\Delta t}A_{x,l}\left(\xzig\zigspvar\right)\right)\right|_{\tau=0}^{\tau=1}-\nabla\sum_{I}W_{\sigma_{0I}}\left(\bfx_{sp,l}\right)\phi_{I,l}+\nonumber \\
 &  & \DDELTATA{\bfx_{sp,l}}\cdot\int_{0}^{1}\rmd\tau\left[\begin{array}{c}
0,\\
\partial_{x}A_{y,l-1}\left(\yzig\zigspmvar\right),\\
\partial_{x}A_{z,l-1}\left(\zzig\zigspmvar\right)
\end{array}\right]\nonumber \\
 &  & =\frac{1}{\Delta t}\left(A_{x,l-1}\left(x_{sp,l},y_{sp,l},z_{sp,l}\right)-A_{x,l}\left(x_{sp,l},y_{sp,l},z_{sp,l}\right)\right)-\nabla\sum_{I}W_{\sigma_{0I}}\left(\bfx_{sp,l}\right)\phi_{I,l}+\nonumber \\
 &  & \DDELTATA{\bfx_{sp,l}}\cdot\int_{0}^{1}\rmd\tau\left[\begin{array}{c}
0,\\
\partial_{x}A_{y,l-1}\left(\yzig\zigspmvar\right)-\partial_{y}A_{x,l-1}\left(\yzig\zigspmvar\right),\\
\partial_{x}A_{z,l-1}\left(\zzig\zigspmvar\right)-\partial_{z}A_{x,l-1}\left(\zzig\zigspmvar\right)
\end{array}\right]\nonumber \\
 &  & =E_{x,l}\left(\bfx_{sp,l}\right)+\frac{y_{sp,l}-y_{sp,l-1}}{\Delta t}\int_{0}^{1}\rmd\tau B_{z,l-1}\left(\yzig\zigspmvar\right)-\nonumber \\
 &  & \frac{z_{sp,l}-z_{sp,l-1}}{\Delta t}\int_{0}^{1}\rmd\tau B_{y,l-1}\left(\zzig\zigspmvar\right),\label{EqnODE1}
\end{eqnarray}
where 
\begin{eqnarray}
\bfB_{l}\left(\bfx\right) & = & \sum_{K}\WTWO{\bfx}\sum_{J}\CURLD_{K,J}\bfA_{J,l}=\sum_{K}\WTWO{\bfx}\bfB_{K,l}~,\\
\bfE_{l}\left(\bfx\right) & = & \sum_{J}\WONE{\bfx}\left(\frac{\bfA_{J,l-1}-\bfA_{J,l}}{\Delta t}-\sum_{I}{\GRADD}_{J,I}\phi_{I,l}\right)=\sum_{J}\WONE{\bfx}\bfE_{J,l}~.
\end{eqnarray}
It clear that \EQ{EqnODE1} is explicit with respect to $x_{sp,l+1}$.
Similar results can be also obtained for solving $y_{sp,l+1}$ and
$z_{sp,l+1}$. We thus have a 1st-order symplectic PIC scheme which
solves $\bfx_{sp,l+1},\bfE_{J,l+1},\bfB_{K,l+1}$ from $\bfx_{sp,l},\bfx_{sp,l-1},\bfE_{J,l},\bfB_{K,l}$.
If we let $\dot{\bfx}_{sp,l}=(\bfx_{sp,l+1}-\bfx_{sp,l})/\Delta t$,
it can be shown that this scheme is equivalent to the 1st-order Hamiltonian
splitting scheme \EQ{EqnFSTHS}. Following this procedure, we can
also construct high-order schemes with more accurate discrete Lagrangians.
For example, a 2nd-order symmetric symplectic method can be built
from the following discrete action, 
\begin{eqnarray}
 &  & S_{d2}=\sum_{l=0}^{N_{t}-1}\Delta tL_{\textrm{dzig2}}\left(\bfx_{sp,2l},\bfx_{sp,2l+1},\bfx_{sp,2l+2},\bfA_{J,l},\bfA_{J,l+1},\phi_{I,l};\Delta t\right)~,\label{EqnActS2}\\
 &  & L_{\textrm{dzig2}}\left(\bfx_{sp,2l},\bfx_{sp,2l+1},\bfx_{sp,2l+2},\bfA_{J,l},\bfA_{J,l+1},\phi_{I,l};\Delta t\right)=L_{\textrm{dfield}}+\nonumber \\
 &  & \sum_{sp}q_{s}\left(m_{s}\left(\frac{\bfx_{sp,2l+1}-\bfx_{sp,2l}}{\Delta t}\right)^{2}+\frac{{\bfx}_{sp,2l+1}-{\bfx}_{sp,2l}}{\Delta t}\cdot\int_{0}^{1}\rmd\tau\bfA_{l}\left(\bfxzig\left(\bfx_{sp,2l},\bfx_{sp,2l+1},\tau\right)\right)+\right.\nonumber \\
 &  & m_{s}\left(\frac{\bfx_{sp,2l+2}-\bfx_{sp,2l+1}}{\Delta t}\right)^{2}+\frac{{\bfx}_{sp,2l+2}-{\bfx}_{sp,2l+1}}{\Delta t}\cdot\int_{0}^{1}\rmd\tau\bfA_{l}\left(\bfzzig\left(\bfx_{sp,2l+1},\bfx_{sp,2l+2},\tau\right)\right)-\nonumber \\
 &  & \left.\sum_{I}W_{\sigma_{0I}}\left(\bfx_{sp,2l}\right)\phi_{I,l}\right)~.\label{EqnLagS2}
\end{eqnarray}
where the zigzag path $\bfzzig$ is 
\begin{eqnarray}
\bfzzig\left(\bfx_{1},\bfx_{2},\tau\right) & = & \left[\begin{array}{c}
\left(x_{1},y_{1},z_{1}+\tau\left(z_{2}-z_{1}\right)\right),\\
\left(x_{1},y_{1}+\tau\left(y_{2}-y_{1}\right),z_{2}\right),\\
\left(x_{1}+\tau\left(x_{2}-x_{1}\right),y_{2},z_{2}\right)
\end{array}\right]~.
\end{eqnarray}
The corresponding equation of motion is again obtained by discrete
variation, 
\begin{eqnarray}
\frac{\partial S_{d2}}{\partial\bfx_{sp,l'}} & = & 0,\quad\textrm{for }1\leq l'\leq2N_{t}-1~,\\
\frac{\partial S_{d2}}{\partial\bfA_{J,l}} & = & 0,\quad\textrm{for }1\leq l\leq N_{t}-1~.
\end{eqnarray}
It can be proved that this 2nd-order method is the same as the 2nd-order
Hamiltonian splitting PIC scheme \EQ{EqnHAMS2}.

\subsection{Structure-preserving geometric relativistic symplectic PIC scheme}

In many practical problems, such as runaway electrons in tokamaks,
astrophysical plasmas, and high-power laser plasmas, effects of relativistic
particles are often significant or dominating. Using the variational
formalism, the structure-preserving geometric formalism can be extended
to relativistic plasmas.

The Lagrangian for relativistic charged particles and electromagnetic
fields is 
\begin{eqnarray}
L_{r}=\iiint\rmd\bfx\left(\frac{\epsilon_{0}}{2}\left(-\dot{\bfA}\left(\bfx\right)-\nabla\phi\left(\bfx\right)\right)^{2}-\frac{1}{2\mu_{0}}\left(\nabla\times\bfA\left(\bfx\right)\right)^{2}+\right.\nonumber \\
\left.\sum_{sp}\delta\left(\bfx-\bfx_{sp}\right)\left(-m_{s}\rmc^{2}\sqrt{1-(\dot{\bfx}_{sp}/\rmc)^{2}}+q_{s}\bfA\left(\bfx\right)\cdot\dot{\bfx}_{sp}-q_{s}\phi\left(\bfx\right)\right)\right)~,\label{EqnLagRVM}
\end{eqnarray}
where $m_{s}$ is the rest mass of the particle of species $s$, $\rmc$
is the light speed in the vacuum. The only difference between $L_{r}$
and the non-relativistic Lagrangian $L$ defined in \EQ{EqnLagVM}
is the relativistic mass factor. Most techniques that have been used
to discretize the electromagnetic fields and the interaction between
particles and electromagnetic fields can be applied to the relativistic
case with little modification.

We let $\rmc=\mu_{0}=\epsilon_{0}=1$ to simplify the notation. To
discretize the relativistic mass factor 
\begin{eqnarray}
L_{rp}\left(\bfx_{sp},\dot{\bfx}_{sp}\right) & = & -m_{s}\sqrt{1-\dot{\bfx}_{sp}^{2}}~,
\end{eqnarray}
we can still use the idea of variational integrator. For example,
a 1st-order discretization of this term is 
\begin{eqnarray}
L_{rpd1}\left(\bfx_{sp,l},\bfx_{sp,l+1};\Delta t\right) & = & L_{rp}\left(\bfx_{sp,l},\frac{\bfx_{sp,l+1}-\bfx_{sp,l}}{\Delta t}\right)~.
\end{eqnarray}
And the total Lagrangian and action integral can be discretized by
a method similar to that used in deriving Eqs. (\ref{EqnSZIG}) and
(\ref{EqnSACT}), 
\begin{eqnarray}
 &  & S_{rd}=\sum_{l=0}^{N_{t}-1}\Delta tL_{\textrm{rdxzig}}\left(\bfx_{sp,l},\bfx_{sp,l+1},\bfA_{J,l},\bfA_{J,l+1},\phi_{I,l};\Delta t\right)~,\label{EqnSRACT}\\
 &  & L_{\textrm{rdxzig}}\left(\bfx_{sp,l},\bfx_{sp,l+1},\bfA_{J,l},\bfA_{J,l+1},\phi_{I,l};\Delta t\right)=L_{\textrm{dfield}}+\nonumber \\
 &  & \sum_{sp}\left(L_{rpd1}\left(\bfx_{sp,l},\bfx_{sp,l+1};\Delta t\right)+q_{s}\left(\DDELTATA{{\bfx}_{sp,l+1}}\cdot\int_{0}^{1}\rmd\tau\bfA_{l}\left(\bfxzig\left(\bfx_{sp,l},\bfx_{sp,l+1},\tau\right)\right)\right)-\right.\nonumber \\
 &  & \left.q_{s}\sum_{I}W_{\sigma_{0I}}\left(\bfx_{sp,l}\right)\phi_{I,l}\right)~.\label{EqnRSZIG}
\end{eqnarray}
The final dynamic equations are derived from the discrete variation,
\begin{eqnarray}
\frac{\partial S_{rd}}{\partial\bfx_{sp,l}} & = & 0~,\label{EqnDSRDX}\\
\frac{\partial S_{rd}}{\partial\bfA_{J,l}} & = & 0~,\label{EqnDSRDA}\\
\frac{\partial S_{rd}}{\partial\phi_{I,l}} & = & 0~.\label{EqnDSRDPHI}
\end{eqnarray}
It is found that Eqs. (\ref{EqnDSRDA}) and (\ref{EqnDSRDPHI}) are
the same as Eqs. (\ref{EqnDEDTREL}) and (\ref{EqnDERHOREL}), respectively.
Unfortunately \EQ{EqnDSRDX} is no longer explicit, and a nonlinear
solver, e.g., the Newton iteration method, is needed to find the solution
for every $\bfx_{sp,l+1}$. But the time advance for the fields is
still explicit.

\subsection{Discrete gauge symmetry and discrete charge conservation law}

For continuous Hamiltonian systems, conservation properties have close
connection with Lie group symmetries, as stated by Noether's theorem
\cite{noether1971invariant}. For example, the energy-momentum conservation
is due to the symmetry of space-time. Such connection for discrete
systems have been investigated in the existing literature \cite{lee1987difference,wendlandt1997mechanical,marsden2001discrete,DORODNITSYN2001307,hairer2006geometric,Hydonrspa20110158,Dorodnitsynbook,Hydon14,xiao2017local}.
In this sub-section, we introduce the discrete gauge symmetry and
discrete charge conservation law for our geometric PIC schemes. Squire
et al. \cite{squire2012geometric} first pointed out the connection
between discrete gauge symmetry and discrete charge conservation.
Shadwick et al. developed a geometric spatial discretization \cite{shadwick2014variational}
with gauge invariance for the Vlasov-Poisson system. Charge conservative
geometric electromagnetic PIC methods in cubic meshes are developed
by Xiao et al. \cite{xiao2015explicit}, and it has been applied to
ideal two-fluid systems as well \cite{xiao2016explicit}. FEEC is
also used to construct geometric PIC algorithms with discrete gauge
symmetry and charged conservation law \cite{he2016hamiltonian,kraus2017gempic}.
The charge conservation property can exist in non-Hamiltonian discrete
particle-field systems \cite{buneman1968relativistic,morse1971numerical,eastwood1991virtual,villasenor1992rigorous,esirkepov2001exact,umeda2003new,pinto2014charge,moon2015exact}.

Now we prove the connection between gauge symmetry and charged conservation
law when the system is discretized in both space and time. First,
let's define the meaning of discrete gauge symmetry. The discrete
system is gauge symmetric, or gauge free, if under the transformation
\begin{eqnarray}
\bfA_{J,l}\rightarrow\bfA_{J,l}' & = & \bfA_{J,l}+\sum_{I}{\GRADD}_{J,I}\psi_{I,l}~,\label{eq:gauge-a}\\
\phi_{I,l}\rightarrow\phi_{I,l}' & = & \phi_{I,l}-\frac{\psi_{I,l}-\psi_{I,l-1}}{\Delta t}~,\label{eq:gauge-phi}
\end{eqnarray}
the Lagrangian of the system changes at most by a total time difference
term. Here, where $\psi_{I,l}$ is an arbitrary discrete 0-form. Note
that this definition of gauge-symmetry is for system that are discrete
in both space and time, which is different from the definition \eqref{eq:gs1}
and \eqref{eq:gs2} for the systems that are discrete in space but
continuous in time. 

For discrete Lagrangians of the particle-field system defined by Eqs.~(\ref{EqnSZIG}),
(\ref{EqnLagS2}) and (\ref{EqnRSZIG}), we can prove that they are
gauge symmetric. As an example, Let's consider the Lagrangian $L_{\textrm{dxzig}}$
specified by Eq.\,(\ref{EqnSZIG}), which contains the following
electromagnetic field term $L_{\textrm{dfield}}$ and the interaction
term $L_{\textrm{int1}}$ , 
\begin{eqnarray}
L_{\textrm{dfield}} & = & \frac{1}{2}\sum_{J}\left(-\DDELTATA{\bfA_{J,l+1}}-\sum_{I}{\GRADD}_{J,I}\phi_{I,l+1}\right)^{2}\Delta V-\nonumber \\
 &  & \frac{1}{2}\sum_{K}\left(\sum_{J}\CURLD_{K,J}\bfA_{J,l}\right)^{2}\Delta V~,\\
L_{\textrm{int1}} & = & \sum_{sp}q_{s}\DDELTATA{{\bfx}_{sp,l+1}}\cdot\int_{0}^{1}\rmd\tau\bfA_{l}\left(\bfxzig\left(\bfx_{sp,l},\bfx_{sp,l+1},\tau\right)\right)-\nonumber \\
 &  & \sum_{sp}q_{s}\sum_{I}W_{\sigma_{0I}}\left(\bfx_{sp,l}\right)\phi_{I,l}~.
\end{eqnarray}
When the discrete field $\bfA_{J,l}$ and $\phi_{I,l}$ are transformed
according to Eqs.\,\eqref{eq:gauge-a} and \eqref{eq:gauge-phi},
only $L_{\textrm{dfield}}$ and $L_{\textrm{int1}}$ are affected,
and total Lagrangian changes into 
\begin{eqnarray}
 &  & L_{\textrm{dxzig}}'=L_{\textrm{dxzig}}+\DDELTAT{F_{1}\left(l\right)}~,\\
 &  & F_{1}\left(l\right)=\sum_{sp,I}q_{s}\psi_{I,l}\WZERO{\bfx_{sp,l}}~.
\end{eqnarray}
This shows that $L_{\textrm{dxzig}}$ is gauge symmetric. In this
case, the discrete action integral only changes by a boundary term,
\begin{eqnarray}
S_{\textrm{dzig}}'=S_{\textrm{dzig}}+F_{1}\left(N_{t}\right)-F_{1}\left(0\right)~.\label{EqnSDZIGFINAL}
\end{eqnarray}
And discrete dynamic equations do not change under the gauge transform.

In fact, this gauge symmetry implies the charge conservation. The
proof is straightforward due to the structure-preserving nature of
the space-time discretization. According to Eq. (\ref{EqnSDZIGFINAL}),
the partial derivative of $S_{\textrm{dzig}}'$ with respect to $\psi_{I,l}$
should be zero when $l\neq0$ and $l\neq N_{t}$, which yields

\begin{equation}
\sum_{sp}q_{s}\left(\DDELTATA{{\bfx}_{sp,l+1}}\cdot\int_{0}^{1}\rmd\tau\sum_{J}{\GRADD}_{J,I}\WONE{\bfxzig\left(\bfx_{sp,l},\bfx_{sp,l+1},\tau\right)}-\DDELTAT{W_{\sigma_{0I}}\left(\bfx_{sp,l}\right)}\right)=0~.\label{EqnDCGT}
\end{equation}
Inserting the $\rho_{I,l}$ and $\bfJ_{J,l}$ defined in Eqs.~(\ref{EqnRHODEF})
and (\ref{EqnCURRDEF}) into Eq. (\ref{EqnDCGT}) leads to the discrete
charge conservation law, 
\begin{eqnarray}
\sum_{J}\GRADD_{J,I}\bfJ_{J,l}-\DDELTAT{\rho_{I,l}} & = & 0~,
\end{eqnarray}
i.e., 
\begin{eqnarray}
\sum_{J}\DIVDP_{I,J}\bfJ_{J,l}+\DDELTAT{\rho_{I,l}} & = & 0~,
\end{eqnarray}
where $\DIVDP$ is the transpose of $-\GRADD$ defined in \EQ{EqnDEFDIVDP}.

If \EQ{EqnRHO} is satisfied initially, according to \EQ{EqnE},
\begin{eqnarray}
-\sum_{J}\nabla_{dJ,I}\DDELTAT{\bfE_{J,l}} & = & -\sum_{J}\DIVDP_{I,J}\bfJ_{J,l}=\DDELTAT{\rho_{I,l}}~,
\end{eqnarray}
we can see that \EQ{EqnRHO} will be automatically satisfied for
all time steps and there is no need to solve it.

Similarly, we can prove that the discrete Lagrangians of the particle-field
systems defined by Eqs.\,(\ref{EqnLagS2}) and (\ref{EqnRSZIG})
also admit discrete charge conservation law due to the fact that they
are gauge symmetric.

\subsection{Numeric examples}

Using the \textsl{C programming language} and the \textsl{Message
Passing Interface} (MPI), we have implemented the 2nd-order explicit
Hamiltonian splitting PIC (EHSPIC) algorithm, the 1st-order variational
symplectic charge-conservative relativistic PIC (VSCRPIC) algorithm,
and the conventional Boris-Yee PIC (BYPIC) scheme for comparison study.
Among these PIC schemes, the VSCRPIC algorithm is relativistic, and
both EHSPIC and VSCRPIC algorithms are structure-preserving and geometric.
Two numerical examples are given to test and compare these algorithms.

\begin{figure}
\subfloat[]{\includegraphics[width=0.49\textwidth]{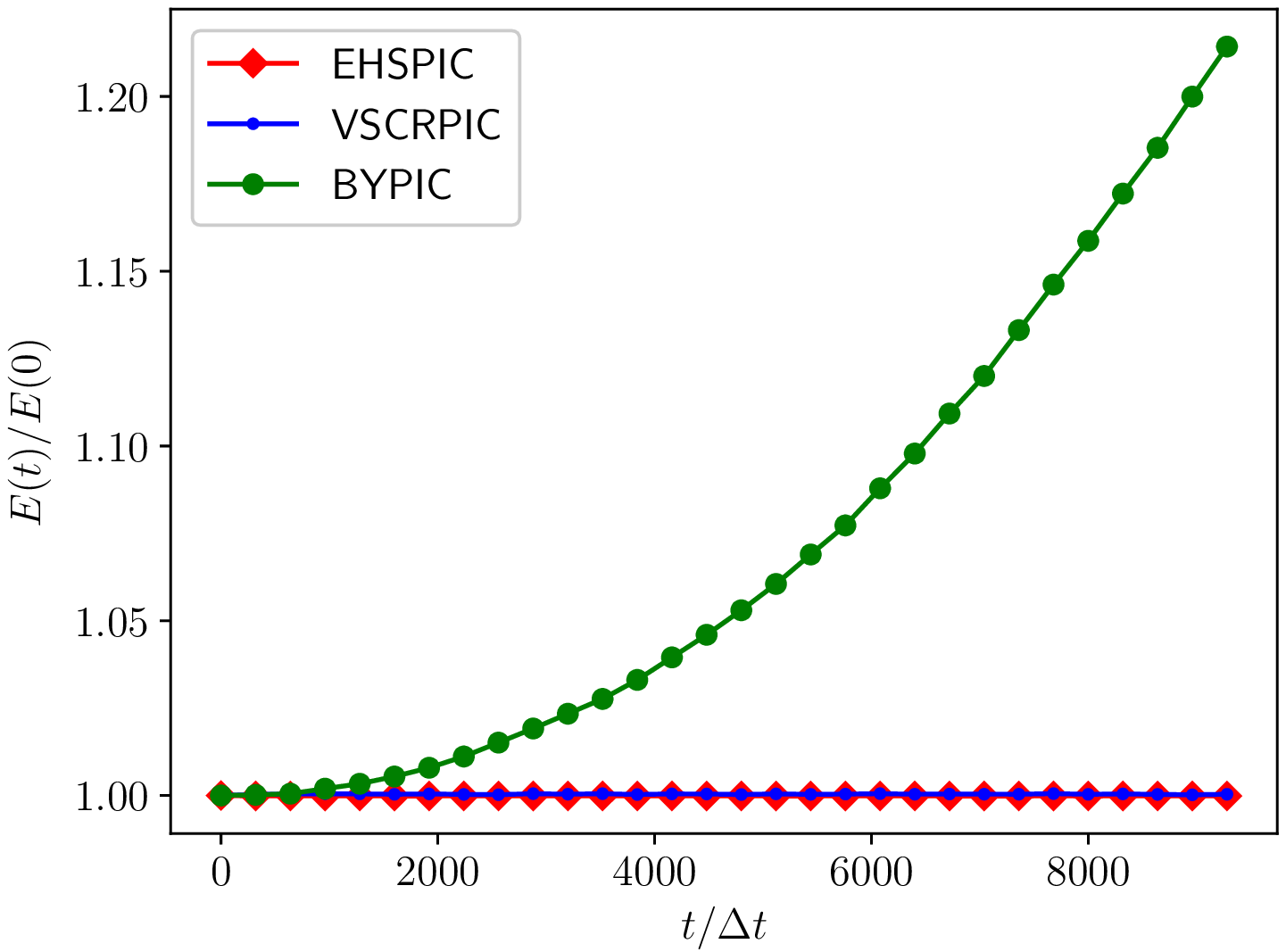}

}\subfloat[]{\includegraphics[width=0.49\textwidth]{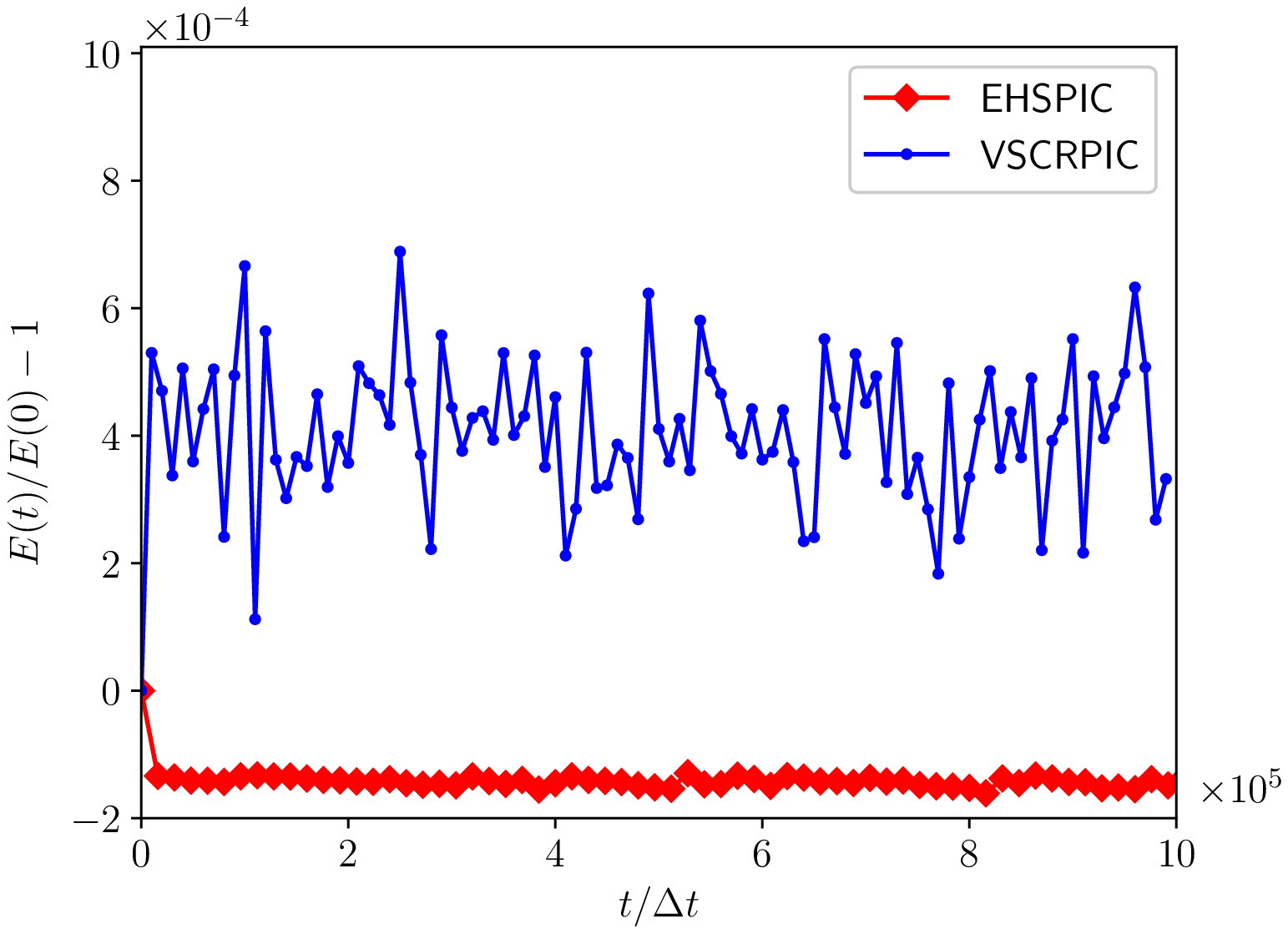}

}

\caption{The evolution of total energy calculated by the conventional BYPIC
scheme and the structure-preserving VSCRPIC and EHSPIC algorithms
in $10^{4}$ time steps (a), and the total energy calculated by two
geometric PIC methods in $10^{6}$ time steps. }

\label{FigEneall} 
\end{figure}

The first example tests the long-term energy conservation property.
Simulation parameters are chosen as follows, $n_{e}=1.0\EXP{19}\mathrm{m}^{-3}~,q_{e}=1.6\EXP{-19}\mathrm{C}~,m_{e}=9.1\EXP{-31}\mathrm{kg}~,v_{T}=0.02\mathrm{c}~,\bfE_{0}=\bfB_{0}=0~,\Delta x=1.0\EXP{-3}\mathrm{m}~,\Delta t=\Delta x/(2\rmc)~,N_{x}=N_{y}=16~,$
and $N_{z}=6~.$ Here, $n_{e}$, $q_{e}$, $m_{e}$ and $v_{T}$ are
density, charge, mass and thermal velocity of electrons, which are
uniformly distributed in the space and their velocity distribution
is Maxwellian. The electromagnetic fields at the initial time are
$\bfE_{0}$ and $\bfB_{0}$, $\Delta x$ and $\Delta t$ are the grid
size and time step, $\rmc$ is the speed of light in the vacuum, and
$N_{x}$, $N_{y}$ and $N_{z}$ are number of grids at $x$, $y$,
$z$ directions. Periodic boundary conditions are adopted in all three
directions. For each grid cell we put 8 sample particles initially.
For these parameters, the plasma frequency is $\omega_{pe}\sim0.3/\Delta t$.
The time history of the total energy obtained in the simulations is
plotted in \FIG{FigEneall}. We can clearly see that for BYPIC,
the total energy grew significantly after only $10^{4}$ time steps,
which is the well-known numerical heating for conventional PIC methods
\cite{birdsall1991plasma,hockney1988computer,ueda1994study}. This
numerical difficulty is overcome by the structure-preserving geometric
PIC schemes. Even after $10^{6}$ time steps (about $1.67\mu$s or
$3\times10^{5}/\omega_{pe}$) the error on total energy for EHSPIC
and VSCRPIC algorithms is still bounded within a small value.

\begin{figure}
\subfloat[VSCRPIC]{\includegraphics[width=0.49\textwidth]{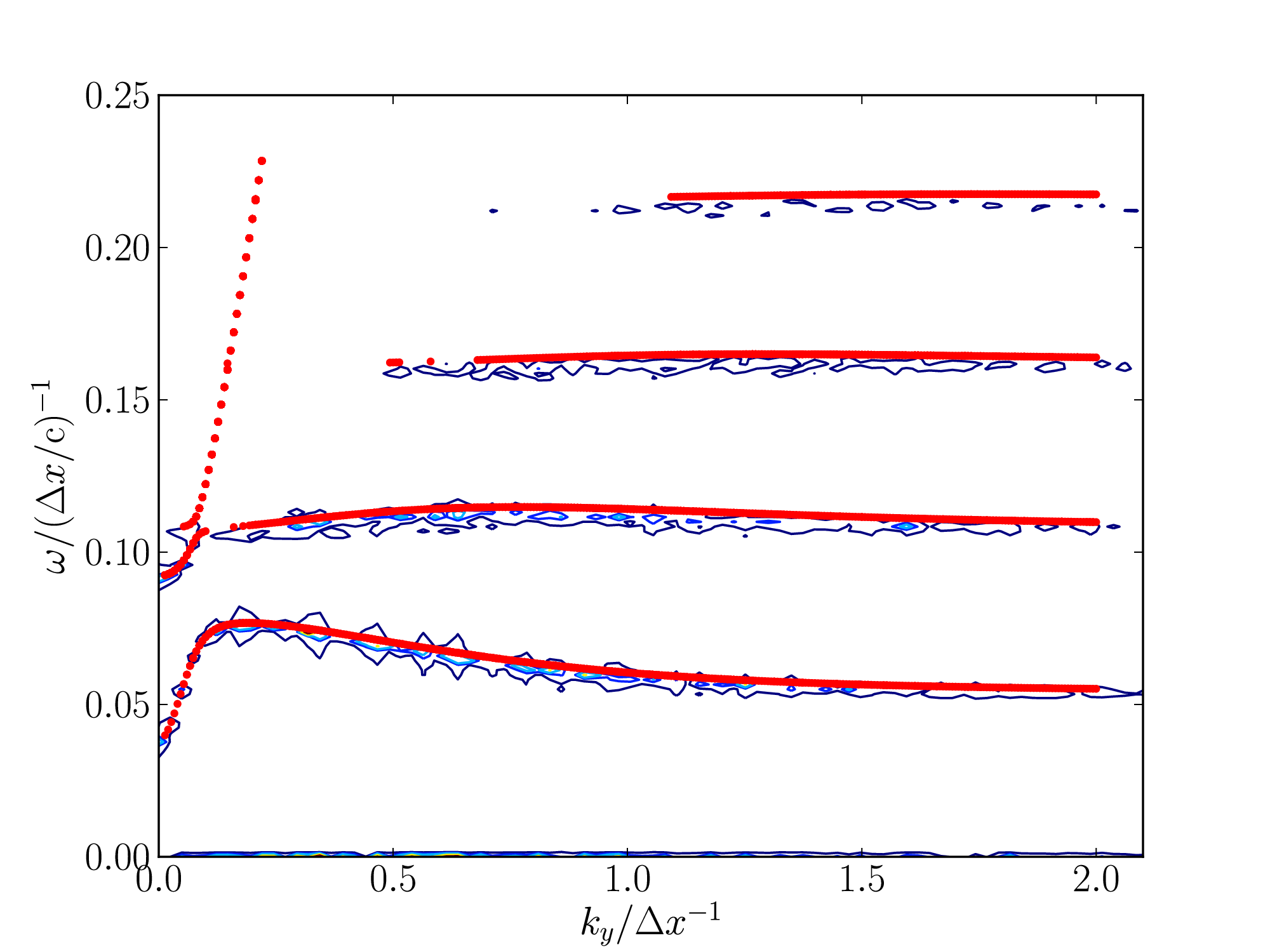}\label{FigDSPREL}

}\subfloat[EHSPIC]{\includegraphics[width=0.49\textwidth]{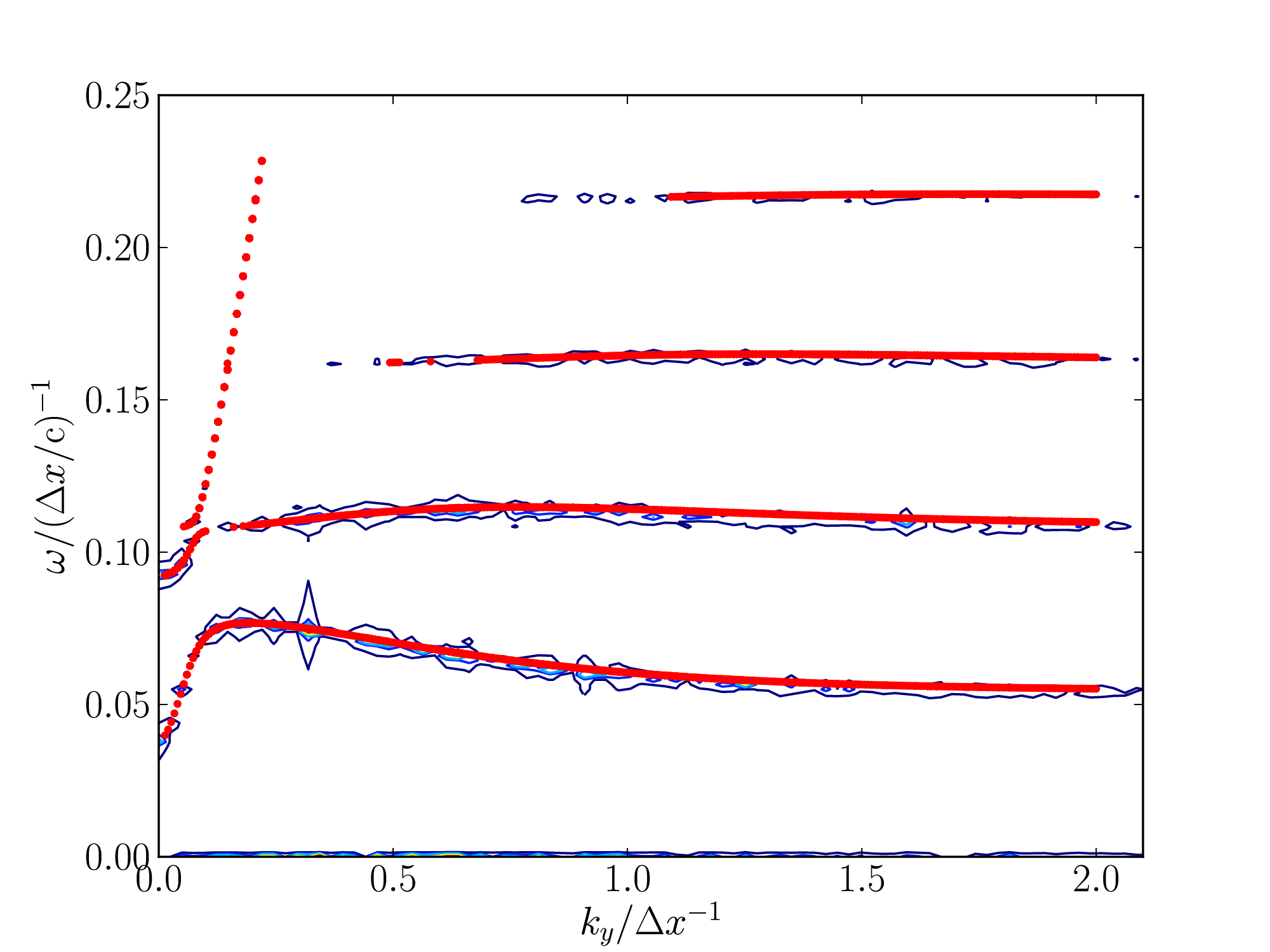}\label{FigDSPES}

}

\subfloat[BYPIC]{\includegraphics[width=0.49\textwidth]{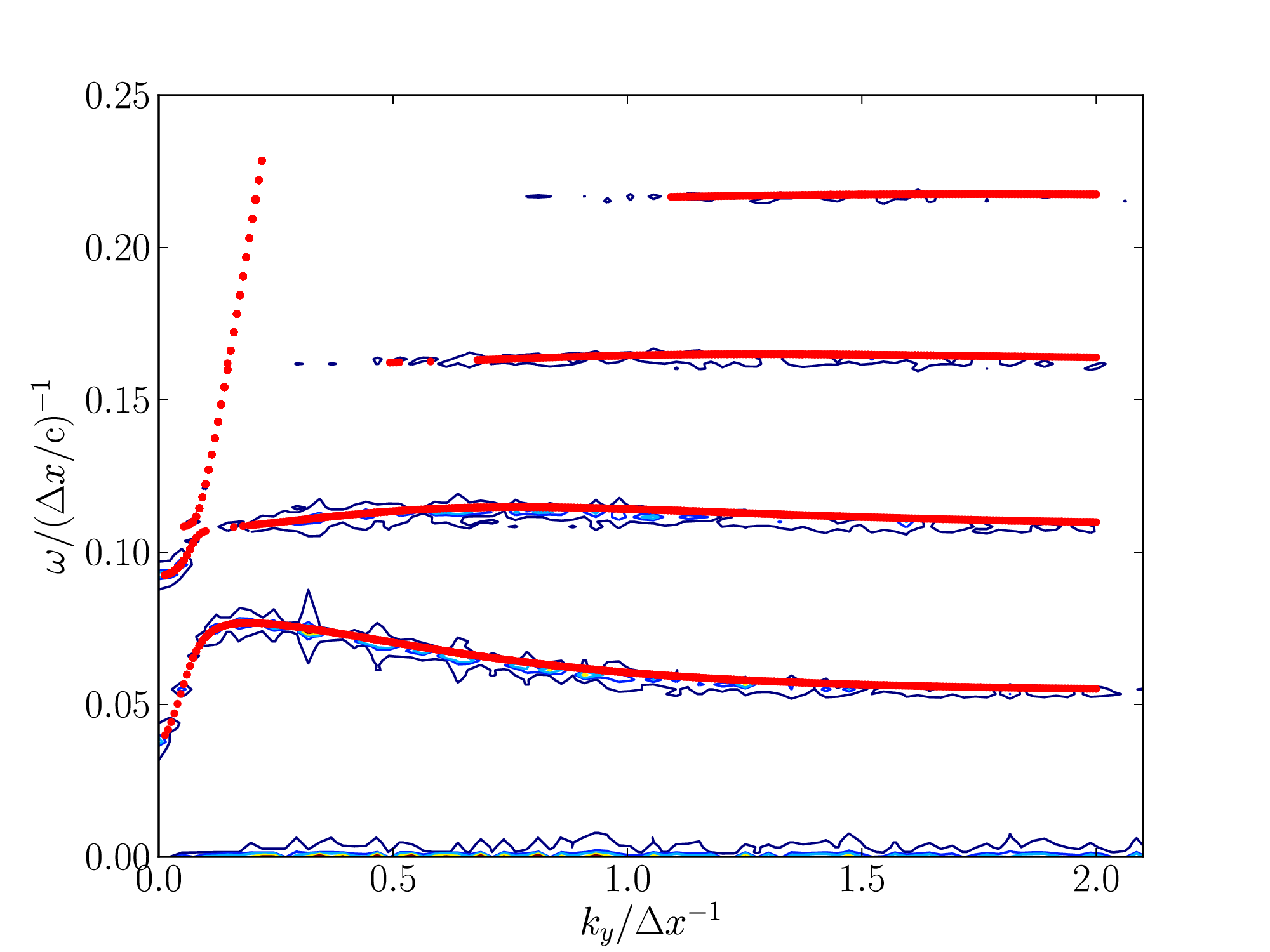}\label{FigDSPBY}

}\subfloat[Total energy]{\includegraphics[width=0.49\textwidth]{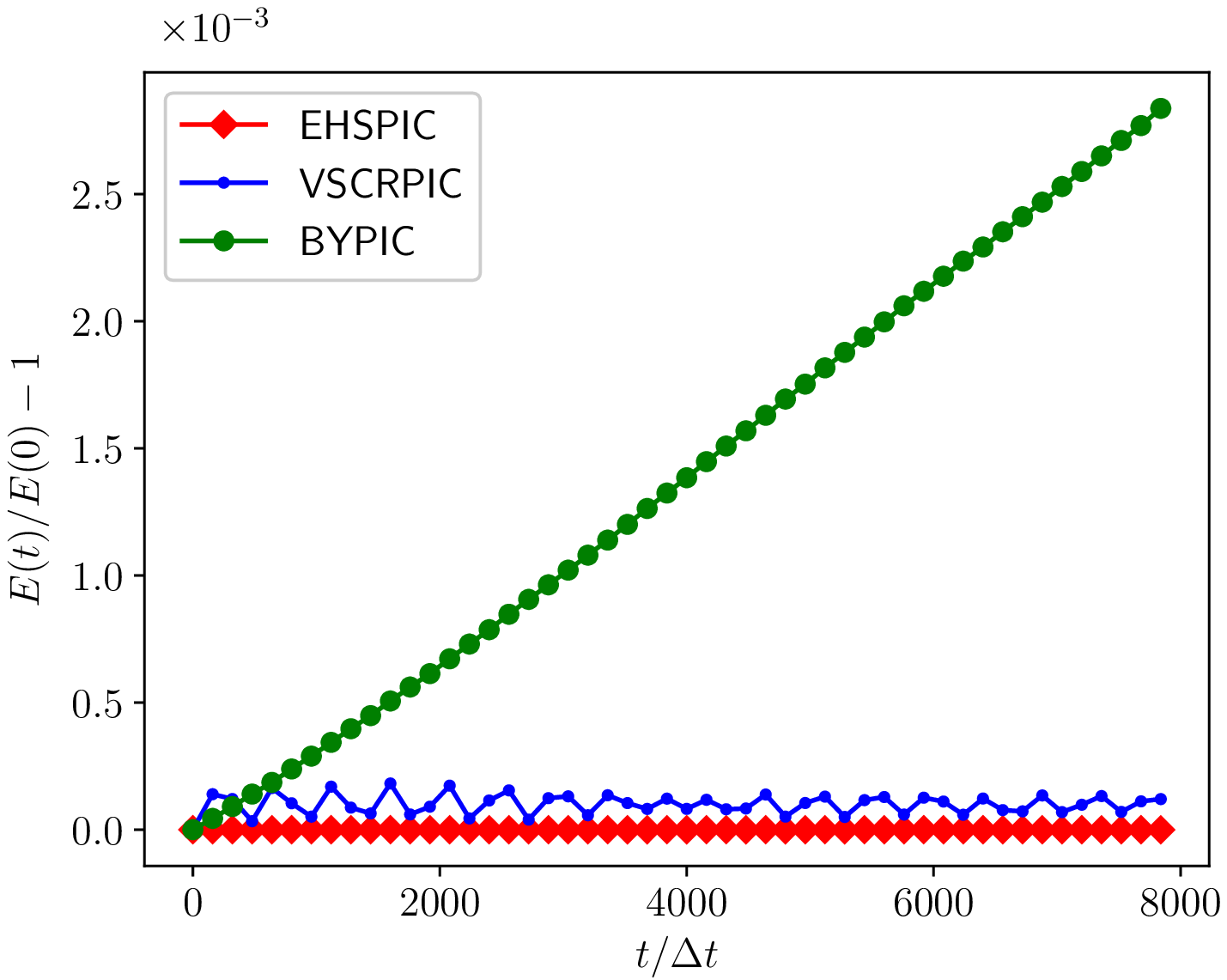}\label{FigXENE}

}

\caption{Space-time spectra of electric fields in the $\bfe_{y}$ direction
of a magnetized hot plasma calculated by the VSCRPIC (a), EHSPIC (b)
and BYPIC (c) methods, and time history of energy error for the three
PIC schemes (d). Red dots are obtained from the analytical dispersion
relation of the corresponding non-relativistic hot magnetized plasma.}

\label{FigSTS} 
\end{figure}

The second example is the dispersion relation of a hot magnetized
plasma. Simulation parameters are chosen as $n_{e}=2.5\EXP{20}\mathrm{m}^{-3}~,q_{e}=1.6\EXP{-19}\mathrm{C}~,m_{e}=9.1\EXP{-31}\mathrm{kg}~,v_{T}=0.07\mathrm{c}~,\bfE_{0}=0~,\bfB_{0}=B_{0}\bfe_{x}~,B_{0}=4.6\mathrm{T}~,\Delta x=2.0\EXP{-5}\mathrm{m}~,\Delta t=\Delta x/(2\rmc)~,N_{x}=N_{z}=1~,$
and $N_{y}=256~.$ Periodic boundaries are used in all 3 directions.
Initially we put 200 sample particles in each grid cell. Due to the
randomness of particle velocity, electromagnetic perturbation will
be excited, and its evolution should satisfy the dispersion relation
of a hot magnetized plasma \cite{stix1992waves}. The space-time spectra
of the electric fields in the $\bfe_{y}$ direction and time history
of the total energy are shown in \FIG{FigSTS}. These results show
that all three methods can recover correctly the analytical dispersion
relation. However, the conventional BYPIC method carries a larger
low frequency noise in the space-time spectrum (Fig.\,\ref{FigDSPBY}),
which is also reflected by the secular growth of the energy error
(Fig.\,\ref{FigXENE}). As expected, the energy error for the two
structure-preserving geometric methods is bounded by a small number
for all simulation time steps. For the VSCRPIC method, the simulated
cyclotron frequency of electrons is slightly lower than the non-relativistic
analytical value due to the relativistic mass factor.

\section{conclusion}

PIC algorithms with different types of discretization will lead to
very different numerical behaviors. Conventional PIC methods are constructed
by discretizing the underpinning integro-differential equations directly,
and the schemes in general do not preserve the geometric structures
of the Vlasov-Maxwell system. As a consequence, numerical errors accumulate
coherently with time and long-term simulation results are not reliable.
On the other hand, structure-preserving geometric PIC algorithms developed
recently utilize modern mathematical techniques, such as a discrete
manifold, interpolating differential forms, and non-canonical symplectic
integrators, to ensure gauge symmetry, space-time symmetry and conservation
of charge, energy-momentum, and the symplectic structure. These highly
desired properties are extremely difficult to achieve using the conventional
PIC schemes. The long-term accuracy and fidelity of these geometric
algorithms have been demonstrated, and it is our vision that structure-preserving
geometric PIC algorithms are most suited for utilizing the upcoming
exascale computing power to simulate the complex behavior of laboratory
and astrophysical plasmas.

\appendix

\section{Discrete operators and interpolating forms}

For easy reference, we list here all definitions and identities of
discrete operators and interpolating forms that are used in this paper.
Most of these can be found in Refs. \cite{xiao2015explicit,xiao2016explicit,xiao2017local}.

Let us begin with discrete difference operators, 
\begin{eqnarray}
\left({\nabla_{\mathrm{d}}}\phi\right)_{i,j,k} & = & [\phi_{i+1,j,k}-\phi_{i,j,k},\phi_{i,j+1,k}-\phi_{i,j,k},\phi_{i,j,k+1}-\phi_{i,j,k}]~.\label{EqnDEFGRADD}\\
\left(\CURLD\bfA\right)_{i,j,k} & = & \left[\begin{array}{c}
\left({A_{z}}_{i,j+1,k}-{A_{z}}_{i,j,k}\right)-\left({A_{y}}_{i,j,k+1}-{A_{y}}_{i,j,k}\right)\\
\left({A_{x}}_{i,j,k+1}-{A_{x}}_{i,j,k}\right)-\left({A_{z}}_{i+1,j,k}-{A_{z}}_{i,j,k}\right)\\
\left({A_{y}}_{i+1,j,k}-{A_{y}}_{i,j,k}\right)-\left({A_{x}}_{i,j+1,k}-{A_{x}}_{i,j,k}\right)
\end{array}\right]^{T}~,\label{EqnDEFcurld}\\
\left({\DIVD}\bfB\right)_{i,j,k} & = & \left({B_{x}}_{i+1,j,k}-{B_{x}}_{i,j,k}\right)+\left({B_{y}}_{i,j+1,k}-{B_{y}}_{i,j,k}\right)+\nonumber \\
 &  & \left({B_{z}}_{i,j,k+1}-{B_{z}}_{i,j,k}\right)~.\label{EqnDEFDIVD}
\end{eqnarray}
These discrete difference operators are linear, so we also use matrices
to denote them. For example, $\GRADD_{J,I}$ is a $3N\times N$ matrix
which maps a discrete scalar field (or 0-form) into a discrete vector
field (or 1-form). Here, $N=N_{x}\times N_{y}\times N_{z}.$

2nd-order Whitney interpolating maps are 
\begin{eqnarray}
\sum_{i,j,k}W_{\sigma_{0,i,j,k}}\left(\bfx\right)\phi_{i,j,k} & \equiv & \sum_{i,j,k}\phi_{i,j,k}W_{1}\left(x\right)W_{1}\left(y\right)W_{1}\left(z\right),\label{EqnDEFW0}\\
\sum_{i,j,k}W_{\sigma_{1,i,j,k}}\left(\bfx\right)\bfA_{i,j,k} & \equiv & \sum_{i,j,k}\left[\begin{array}{c}
{A_{x}}_{i,j,k}W_{1}^{(2)}(x-i)W_{1}(y-j)W_{1}(z-k)\\
{A_{y}}_{i,j,k}W_{1}(x-i)W_{1}^{(2)}(y-j)W_{1}(z-k)\\
{A_{z}}_{i,j,k}W_{1}(x-i)W_{1}(y-j)W_{1}^{(2)}(z-k)~
\end{array}\right]^{T},\label{EqnDEFW1}\\
\sum_{i,j,k}W_{\sigma_{2,i,j,k}}\left(\bfx\right)\bfB_{i,j,k} & \equiv & \sum_{i,j,k}\left[\begin{array}{c}
{B_{x}}_{i,j,k}W_{1}(x-i)W_{1}^{(2)}(y-j)W_{1}^{(2)}(z-k)\\
{B_{y}}_{i,j,k}W_{1}^{(2)}(x-i)W_{1}(y-j)W_{1}^{(2)}(z-k)\\
{B_{z}}_{i,j,k}W_{1}^{(2)}(x-i)W_{1}^{(2)}(y-j)W_{1}(z-k)
\end{array}\right]^{T},\label{EqnDEFW2}\\
\sum_{i,j,k}W_{\sigma_{3,i,j,k}}\left(\bfx\right)\rho_{i,j,k} & \equiv & \sum_{i,j,k}\rho_{i,j,k}W_{1}^{(2)}(x-i)W_{1}^{(2)}(y-j)W_{1}^{(2)}(z-k),\label{EqnDEFW3}
\end{eqnarray}
\begin{eqnarray}
W_{1}^{(2)}\left(x\right) & = & -\left\{ \begin{array}{lc}
W_{1}'\left(x\right)+W_{1}'\left(x+1\right)+W_{1}'\left(x+2\right)~, & -1\leq x<2~,\\
0~, & \textrm{otherwise}~.
\end{array}\right.
\end{eqnarray}
Here, $W_{1}$ is a $C^{2}$ one-dimensional interpolating function
that is only non-zero in the interval $(-2,2)$, and it can be chosen
to be the $W_{1}$ defined by \EQ{EqnW1}.

We have also used the following backward difference operators, 
\begin{eqnarray}
\left({\nabla_{\mathrm{d}}^{*}}\phi\right)_{i,j,k} & = & [\phi_{i,j,k}-\phi_{i-1,j,k},\phi_{i,j,k}-\phi_{i,j-1,k},\phi_{i,j,k}-\phi_{i,j,k-1}]~.\label{EqnDEFGRADDP}\\
\left(\CURLDP\bfA\right)_{i,j,k} & = & \left[\begin{array}{c}
\left({A_{z}}_{i,j,k}-{A_{z}}_{i,j-1,k}\right)-\left({A_{y}}_{i,j,k}-{A_{y}}_{i,j,k-1}\right)\\
\left({A_{x}}_{i,j,k}-{A_{x}}_{i,j,k-1}\right)-\left({A_{z}}_{i,j,k}-{A_{z}}_{i-1,j,k}\right)\\
\left({A_{y}}_{i,j,k}-{A_{y}}_{i-1,j,k}\right)-\left({A_{x}}_{i,j,k}-{A_{x}}_{i,j-1,k}\right)
\end{array}\right]^{T}~,\label{EqnDEFcurldp}\\
\left({\DIVDP}\bfB\right)_{i,j,k} & = & \left({B_{x}}_{i,j,k}-{B_{x}}_{i-1,j,k}\right)+\left({B_{y}}_{i,j,k}-{B_{y}}_{i,j-1,k}\right)+\nonumber \\
 &  & \left({B_{z}}_{i,j,k}-{B_{z}}_{i,j,k-1}\right)~.\label{EqnDEFDIVDP}
\end{eqnarray}
These operators have close connections with the corresponding forward
difference operators, 
\begin{eqnarray}
\nabla_{\mathrm{d}J,I} & = & -\DIVDP_{I,J}~,\\
\CURLD_{K,J} & = & \CURLDP_{J,K}~,\\
\DIVD_{L,K} & = & -\nabla_{\mathrm{d}K,L}^{*}~.
\end{eqnarray}
It can be easily checked that these discrete operators and Whitney
interpolating maps satisfy following identities,
\begin{eqnarray}
\sum_{J,I}\CURLD_{K,J}{\nabla_{\rmd}}_{J,I}\phi_{I} & = & 0~,\label{EqnCURLDGRADDZERO1}\\
\sum_{K,J}\DIVD_{L,K}\CURLD_{K,J}\bfA_{J} & = & 0~,\label{EqnDIVDCURLDZERO1}\\
\sum_{K,L}\CURLDP_{J,K}{\nabla_{\rmd}^{*}}_{K,L}\phi_{L} & = & 0~,\label{EqnCURLDPGRADDPZERO1}\\
\sum_{J,K}\DIVDP_{I,J}\CURLDP_{J,K}\bfA_{K} & = & 0~.\label{EqnDIVDPCURLDPZERO1}
\end{eqnarray}
\begin{eqnarray}
\nabla\sum_{I}\WZERO{\bfx}\phi_{I} & = & \sum_{I,J}\WONE{\bfx}{\nabla_{\mathrm{d}}}_{J,I}\phi_{I}~,\\
\nabla\times\sum_{J}\WONE{\bfx}\bfA_{J} & = & \sum_{J,K}\WTWO{\bfx}{\CURLD}_{K,J}\bfA_{J}~.\label{EqnD1to2FORM1}\\
\nabla\times\sum_{K}\WTWO{\bfx}\bfB_{K} & = & \sum_{K,L}\WTHREE{\bfx}{\DIVD}_{L,K}\bfB_{J}~.\label{EqnD2to3FORM}
\end{eqnarray}

\section{Iteration schemes for PIC methods}

In this appendix, we list the three PIC iteration schemes that are
used in this work, which are the Boris-Yee PIC (BYPIC) algorithm ,
the explicit Hamiltonian splitting PIC (EHSPIC) algorithm and the
variational symplectic charge-conservative relativistic PIC (VSCRPIC)
algorithm. 

\subsection{BYPIC scheme}

In the BYPIC scheme, electromagnetic fields are solved by using the
Yee-FDTD scheme, i.e., 
\begin{eqnarray}
\bfB_{K,l+1/2} & = & \bfB_{K,l-1/2}-\Delta t\sum_{J}\CURLD_{K,J}\bfE_{K,l}~,\\
\bfE_{J,l+1} & = & \bfE_{J,l}+\Delta t\left(\sum_{J}\CURLDP_{J,K}\bfB_{K,l+1/2}-\bfJ_{J,l+1/2}\right)~.
\end{eqnarray}
Positions and velocities of particles are updated by the Boris algorithm,
i.e., 
\begin{eqnarray}
\frac{\bfv_{sp,l+1/2}-\bfv_{sp,l-1/2}}{\Delta t} & = & \frac{q_{s}}{m_{s}}\left(\bfE_{sp,l}+\frac{\bfv_{sp,l+1}+\bfv_{sp,l}}{2}\times\bfB_{sp,l}\right)~,\\
\frac{\bfx_{sp,l+1}-\bfx_{sp,l}}{\Delta t} & = & \bfv_{sp,l+1/2}~,
\end{eqnarray}
where 
\begin{eqnarray}
\bfE_{sp,l} & = & \sum_{J}\bfE_{J,l}W\left(\bfx_{sp,l}-\bfx_{J}\right)~,\label{EqnInterpE}\\
\bfB_{sp,l} & = & \sum_{J}\left(\bfB_{K,l-1/2}-\frac{\Delta t}{2}\sum_{J}\CURLD_{K,J}\bfE_{K,l}\right)W\left(\bfx_{sp,l}-\bfx_{K}\right)~,\\
\bfJ_{J,l+1/2} & = & \sum_{sp}q_{s}\bfv_{sp,l+1/2}W\left(\bfx_{sp,l}+\frac{\Delta t}{2}\bfv_{sp,l+1/2}-\bfx_{J}\right)~,
\end{eqnarray}
and $W$ is the interpolating function. The initial conditions for
the electromagnetic fields and particles' positions and velocities
are $\bfB_{K,-1/2},\bfE_{J,0},\bfx_{sp,0},\bfv_{sp,-1/2}$. 

\subsection{EHSPIC scheme}

The EHSPIC scheme is a splitting method. As discussed in \SEC{SecEHNCSPIC},
the basic one-step maps for the sub-systems are $\Theta_{E}\left(\Delta t\right),\Theta_{B}\left(\Delta t\right),\Theta_{x}\left(\Delta t\right),\Theta_{y}\left(\Delta t\right),\Theta_{z}\left(\Delta t\right)$,
that map $z=[\bfB_{K},\bfE_{J},\bfx_{sp},\dot{\bfx}_{sp}]$ into new
$z'=[\bfB_{K}',\bfE_{J}',\bfx_{sp}',\dot{\bfx}_{sp}']$. They are
\begin{eqnarray}
\Theta_{E}\left(\Delta t\right) & : & \left\{ \begin{array}{ccl}
\bfE_{J}' & = & \bfE_{J}~,\\
\bfB_{K}' & = & \bfB_{K}-\Delta t\sum_{J}\CURLD_{K,J}\bfE_{J}~,\\
\bfx_{sp}' & = & \bfx_{sp}~,\\
\dot{\bfx}_{sp}' & = & \dot{\bfx}_{sp}+\frac{q_{s}}{m_{s}}\Delta t\sum_{J}W_{\sigma_{1J}}\left(\bfx_{sp}\right)\bfE_{J}~,
\end{array}\right.\\
\Theta_{B}\left(\Delta t\right) & : & \left\{ \begin{array}{ccl}
\bfE_{J}' & = & \bfE_{J}+\Delta t\sum_{K}\CURLD_{K,J}\bfB_{K}~,\\
\bfB_{K}' & = & \bfB_{K}~,\\
\bfx_{sp}' & = & \bfx_{sp}~,\\
\dot{\bfx}_{sp}' & = & \dot{\bfx}_{sp}~.
\end{array}\right.
\end{eqnarray}

\begin{eqnarray}
\Theta_{x}\left(\Delta t\right) & : & \left\{ \begin{array}{ccl}
\bfE_{J}' & = & \bfE_{J}-\int_{0}^{\Delta t}dt'\sum_{s}\frac{q_{s}}{\Delta V}\dot{x}_{sp}(t)\bfe_{x}W_{\sigma_{1J}}\left(\bfx_{sp}+\dot{x}_{sp}t'\bfe_{x}\right)~,\\
\bfB_{K}' & = & \bfB_{K}~,\\
\bfx_{sp}' & = & \bfx_{sp}+\Delta t\dot{x}_{sp}\bfe_{x}~,\\
\dot{\bfx}_{sp}' & = & \dot{\bfx}_{sp}+\frac{q_{s}}{m_{s}}\dot{x}_{sp}\bfe_{x}\times\int_{0}^{\Delta t}dt'\sum_{K}W_{\sigma_{2K}}\left(\bfx_{sp}+\dot{x}_{sp}t'\bfe_{x}\right)\bfB_{K}~.
\end{array}\right.\\
\Theta_{y}\left(\Delta t\right) & : & \left\{ \begin{array}{ccl}
\bfE_{J}' & = & \bfE_{J}-\int_{0}^{\Delta t}dt'\sum_{s}\frac{q_{s}}{\Delta V}\dot{y}_{sp}(t)\bfe_{y}W_{\sigma_{1J}}\left(\bfx_{sp}+\dot{y}_{sp}t'\bfe_{y}\right)~,\\
\bfB_{K}' & = & \bfB_{K}~,\\
\bfx_{sp}' & = & \bfx_{sp}+\Delta t\dot{y}_{sp}\bfe_{y}~,\\
\dot{\bfx}_{sp}' & = & \dot{\bfx}_{sp}+\frac{q_{s}}{m_{s}}\dot{y}_{sp}\bfe_{y}\times\int_{0}^{\Delta t}dt'\sum_{K}W_{\sigma_{2K}}\left(\bfx_{sp}+\dot{y}_{sp}t'\bfe_{y}\right)\bfB_{K}~.
\end{array}\right.\label{EQNTHETAY}\\
\Theta_{z}\left(\Delta t\right) & : & \left\{ \begin{array}{ccl}
\bfE_{J}' & = & \bfE_{J}-\int_{0}^{\Delta t}dt'\sum_{s}\frac{q_{s}}{\Delta V}\dot{z}_{sp}(t)\bfe_{z}W_{\sigma_{1J}}\left(\bfx_{sp}+\dot{z}_{sp}t'\bfe_{z}\right)~,\\
\bfB_{K}' & = & \bfB_{K}~,\\
\bfx_{sp}' & = & \bfx_{sp}+\Delta t\dot{z}_{sp}\bfe_{z}~,\\
\dot{\bfx}_{sp}' & = & \dot{\bfx}_{sp}+\frac{q_{s}}{m_{s}}\dot{z}_{sp}\bfe_{z}\times\int_{0}^{\Delta t}dt'\sum_{K}W_{\sigma_{2K}}\left(\bfx_{sp}+\dot{z}_{sp}t'\bfe_{z}\right)\bfB_{K}~.
\end{array}\right.\label{EQNTHETAZ}
\end{eqnarray}
If we use $\Theta_{f}\Theta_{g}$ to denote a map that is composed
by $\Theta_{f}$ and $\Theta_{g}$, 
\begin{eqnarray}
(\Theta_{f}\Theta_{g})(z)=\Theta_{f}(\Theta_{g}(z))~,
\end{eqnarray}
then the 1st-, 2nd- and $2(l+1)$-th order methods $\Theta_{1}\left(\Delta t\right),\Theta_{2}\left(\Delta t\right),\Theta_{2(l+1)}\left(\Delta t\right)$
can be constructed as 
\begin{eqnarray}
\Theta_{1}\left(\Delta t\right) & = & \Theta_{x}\left(\Delta t\right)\Theta_{y}\left(\Delta t\right)\Theta_{z}\left(\Delta t\right)\Theta_{E}\left(\Delta t\right)\Theta_{B}\left(\Delta t\right)~,\\
\Theta_{2}\left(\Delta t\right) & = & \Theta_{E}\left(\Delta t/2\right)\Theta_{x}\left(\Delta t/2\right)\Theta_{y}\left(\Delta t/2\right)\Theta_{z}\left(\Delta t/2\right)\Theta_{B}\left(\Delta t\right)\nonumber \\
 &  & \Theta_{z}\left(\Delta t/2\right)\Theta_{y}\left(\Delta t/2\right)\Theta_{x}\left(\Delta t/2\right)\Theta_{E}\left(\Delta t/2\right)~,\\
\Theta_{2(l+1)}(\Delta t) & = & \Theta_{2l}(\alpha_{l}\Delta t)\Theta_{2l}(\beta_{l}\Delta t)\Theta_{2l}(\alpha_{l}\Delta t)~,\\
\alpha_{l} & = & 1/(2-2^{1/(2l+1)})~,\nonumber \\
\beta_{l} & = & 1-2\alpha_{l}~.\nonumber 
\end{eqnarray}

\subsection{VSCRPIC scheme}

For the 1st-order VSCRPIC scheme, the algorithm for the electromagnetic
fields is, 
\begin{eqnarray}
\frac{\bfB_{K,l}-\bfB_{K,l-1}}{\Delta t} & = & -\sum_{J}\CURLD_{K,J}\bfE_{J,l}~,\\
\frac{\bfE_{J,l+1}-\bfE_{J,l}}{\Delta t} & = & \sum_{K}\CURLDP_{J,K}\bfB_{K,l}-\bfJ_{J,l}~,
\end{eqnarray}
and the algorithm for the dynamics of the $sp$-th particle is 
\begin{equation}
\frac{m_{s}}{\Delta t}\left(\frac{\bfv_{sp,l}}{\sqrt{1-\bfv_{sp,l}}}-\frac{\bfv_{sp,l-1}}{\sqrt{1-\bfv_{sp,l-1}}}\right)=q_{s}\left(\bfE_{sp,l}+\bfv_{sp,l-1}\cdot\hat{\bfB}_{sp,l-1}+\bfv_{sp,l}\cdot\hat{\bfB}_{sp,l}^{*}\right)~,
\end{equation}
where 
\begin{equation}
\bfE_{sp,l}=\sum_{J}\bfE_{J,l}\WONE{\bfx_{sp,l}}~,
\end{equation}
\begin{equation}
\bfv_{sp,l-1}\cdot\hat{\bfB}_{sp,l-1}=\left[\begin{array}{c}
v_{y,sp,l-1}\int_{0}^{1}\rmd t'B_{z,l-1}\left(x_{sp,l},y_{sp,l-1}+t'\left(y_{sp,l}-y_{sp,l-1}\right),z_{sp,l-1}\right)-\\
v_{z,sp,l-1}\int_{0}^{1}\rmd t'B_{y,l-1}\left(x_{sp,l},y_{sp,l},z_{sp,l-1}+t'\left(z_{sp,l}-z_{sp,l-1}\right)\right)~,\\
v_{z,sp,l-1}\int_{0}^{1}\rmd t'B_{x,l-1}\left(x_{sp,l},y_{sp,l},z_{sp,l-1}+t'\left(z_{sp,l}-z_{sp,l-1}\right)\right)~,\\
0
\end{array}\right]~,
\end{equation}
\begin{equation}
\bfv_{sp,l}\cdot\hat{\bfB}_{sp,l}^{*}=\left[\begin{array}{c}
0,\\
-v_{x,sp,l}\int_{0}^{1}\rmd t'B_{z,l}\left(x_{sp,l}+t'\left(x_{sp,l+1}-x_{sp,l}\right),y_{sp,l},z_{sp,l}\right)~,\\
v_{x,sp,l}\int_{0}^{1}\rmd t'B_{y,l}\left(x_{sp,l}+t'\left(x_{sp,l+1}-x_{sp,l}\right),y_{sp,l},z_{sp,l}\right)-\\
v_{y,sp,l}\int_{0}^{1}\rmd t'B_{x,l}\left(x_{sp,l+1},y_{sp,l}+t'\left(y_{sp,l+1}-y_{sp,l}\right),z_{sp,l}\right)~
\end{array}\right]~,
\end{equation}
\begin{eqnarray}
\bfv_{sp,l} & = & \frac{\bfx_{sp,l+1}-\bfx_{sp,l}}{\Delta t}~,\\
\left[\begin{array}{c}
B_{x,l}\left(\bfx\right),\\
B_{y,l}\left(\bfx\right),\\
B_{z,l}\left(\bfx\right)
\end{array}\right] & = & \sum_{K}\bfB_{K,l}\WTWO{\bfx}~,
\end{eqnarray}
\begin{eqnarray}
\bfJ_{J,l} & = & \sum_{sp}q_{s}\DDELTATA{\bfx_{sp,l+1}}\cdot\int_{0}^{1}\rmd\tau\WONE{\bfxzig\left(\bfx_{sp,l},\bfx_{sp,l+1},\tau\right)}~.
\end{eqnarray}
\begin{acknowledgments}
This research is supported by National Natural Science Foundation
of China (NSFC-11775219, 11775222, 11505186, 11575185 and 11575186),
National Key Research and Development Program (2016YFA0400600, 2016YFA0400601
and 2016YFA0400602), ITER-China Program (2015GB111003, 2014GB124005)
Chinese Scholar Council (201506340103), China Postdoctoral Science
Foundation (2017LH002), and the GeoAlgorithmic Plasma Simulator (GAPS)
Project. 
\end{acknowledgments}

\bibliographystyle{apsrev4-1}
\bibliography{pst_overview_20180327}

\end{document}